\documentclass[lettersize,journal]{IEEEtran}
\def\BibTeX{{\rm B\kern-.05em{\sc i\kern-.025em b}\kern-.08em
		T\kern-.1667em\lower.7ex\hbox{E}\kern-.125emX}}
\usepackage{cite}
\usepackage{algpseudocode}
\usepackage{graphicx}
\usepackage{amsmath}
\usepackage{amsfonts}
\usepackage{epstopdf}
\usepackage{xcolor}
\usepackage{enumerate}
\usepackage{multirow}
\usepackage{transparent}
\usepackage{subfigure}
\usepackage{float}
\usepackage{amssymb}
\usepackage[]{algorithm2e}

\DeclareMathOperator*{\E}{\mathbb{E}}

\newtheorem{theorem}{Theorem}
\newtheorem{remark}{Remark}

\begin{document}
	\title{Low-Complexity OTFS-Based Over-the-Air Computation Design for Time-Varying Channels}
	\author{Xinyu Huang, \textit{Member, IEEE}, Henrik Hellström, \textit{Member, IEEE}, and Carlo Fischione, \textit{Fellow, IEEE}}
	\maketitle
	\begin{abstract}
		This paper investigates over-the-air computation (AirComp) over multiple-access time-varying channels, where devices with high mobility transmit their sensing data to a fusion center (FC) for averaging. To combat the Doppler shift induced by time-varying channels, each device adopts orthogonal time frequency space (OTFS) modulation. Our objective is minimizing the mean squared error (MSE) for the target function estimation. Due to the multipath time-varying channels, the OTFS-based AirComp not only suffers from noise but also interference. Specifically, we propose three schemes, namely S1, S2, and S3, for the target function estimation. S1 directly estimates the target function under the impacts of noise and interference. S2 mitigates the interference by introducing a zero padding-assisted OTFS. In S3, we propose an iterative algorithm to estimate the function in a matrix form. In the numerical results, we evaluate the performance of S1, S2, and S3 from the perspectives of MSE and computational complexity, and compare them with benchmarks. Specifically, compared to benchmarks, S3 outperforms them with a significantly lower MSE but incurs a higher computational complexity. In contrast, S2 demonstrates a reduction in both MSE and computational complexity. Lastly, S1 shows superior error performance at small SNR and reduced computational complexity.
	\end{abstract}
	\begin{IEEEkeywords}
		Over-the-air computation, orthogonal time frequency space modulation, time-varying channels, high-mobility
	\end{IEEEkeywords}
	\section{Introduction}
	The latest developments in wireless communication and smart device technologies have spurred the widespread adoption of the Internet of Things (IoT), enabling the seamless connection of billions of physical objects to the Internet through ubiquitous sensing and computing capabilities \cite{nguyen20216g}. According to Dell report \cite{delltec}, there will be 41.6 billion IoT devices in 2025, capable of generating 79.4 zettabytes (ZB) of data. Traditional data aggregation methods via orthogonal channels may not be applicable in future IoT networks due to the restricted spectrum resources. Recently, over-the-air computation (AirComp) was proposed as an efficient method to overcome this issue, which leverages the signal superposition property over multiple access channels (MAC) to allow rapid wireless data aggregation \cite{csahin2023survey}. Due to the unique characteristics of AirComp in integrating computation and communication, its application in IoT networks, distributed machine learning, and AI-assisted autonomous driving systems holds significant potential \cite{wang2022edge}.
	
	The concept of AirComp was originaly proposed in \cite{nazer2007computation}. In this paper, the authors considered reliably reconstructing a function of sources over Gaussian MAC, where lattice computational coding is applied for signal transmission. The true significance of this work lies in its compelling discovery that interference can be effectively leveraged to enhance computational processes. In \cite{gastpar2008uncoded}, the authors showed that analog signal transmission without coding but with certain pre-processing can achieve a lower mean squared error (MSE) than \cite{nazer2007computation}. In AirComp, as air acts as a computer, the status of channel is important. Most previous studies, e.g., \cite{yang2020federated,shao2021federated}, considered AirComp over time-invariant channels due to its stable channel state infomation (CSI). In \cite{yang2020federated}, the authors investigated AirComp for fast global model aggregation in federated learning (FL), where the authors assumed a constant channel vector between a fusion center (FC) and each device. The authors in \cite{shao2021federated} considered AirComp with residual channel gain variation and symbol-timing asynchrony among devices. They assumed a flat and slow fading channel, such that CSI remains constant.
	
	The next generation of wireless networks are expected to support a wide range of high-mobility terminals, spanning from autonomous vehicles to unmanned aerial vehicles (UAVs), low-earth-orbit (LEO) satellites, and high-speed trains, which leads to a time-varying double-selectively channel. Some previous studies, e.g., \cite{cao2020optimized,zhu2019broadband,tegin2023federated}, considered AirComp over time-varying channels. In \cite{cao2020optimized}, the authors considered a power control problem of AirComp, where the authors minimized the MSE by jointly optimizing the transmit power at devices and a signal scaling factor at the FC. In \cite{zhu2019broadband}, the authors considered a broadband channel and quantified the latency reduction of applying AirComp in FL with respect to conventional orthogonal multiple-access schemes. In \cite{tegin2023federated}, the authors applied orthogonal frequency-division multiplexing (OFDM) to modulate the transmitted model parameters and examined the effects of time-varying channels on the convergence of FL. Moreover, as pre-processing is usually required for AirComp, \cite{cao2020optimized,zhu2019broadband} assumed that the transmitters (TXs) always know the instantaneous channel gains. They also assumed that function computation and estimation are always performed within the channel coherence time. These assumptions require the RX estimates the instantaneous CSI and broadcast them to all TXs, which leads to a frequent signaling with a large overhead. Furthermore, for fast-varying channels, the channel coherence time can be less than the signaling time, which makes AirComp in \cite{cao2020optimized,zhu2019broadband} potentially impossible. Additionally, the high Doppler shifts can generate severe inter-carrier interference (ICI), which degrades the performance of applying OFDM in \cite{tegin2023federated}.
	
	To reduce the signaling overhead and eliminate the undesired effects introduced by Doppler shifts, orthogonal time-frequency space (OTFS) modulation was recently proposed as a promising technique to combat Doppler effects and time-varying channels \cite{hadani2017orthogonal}. In OTFS, the transmitted symbols are modulated on the delay-Doppler (DD) domain. After a series of signal processing operations, the signal is transmitted through the time-varying channel in the time domain. At the FC side, the target function is still estimated on the DD domain. There are several advantages of OTFS compared with the conventionally applied OFDM. First, channels in the DD domain exhibits compactness and sparsity, which effectively eliminate the ICI induced by the Doppler shift. Second, the data modulated on the DD domain principally experiences the whole fluctions of the channel in the time-frequency (TF) domain, which potentially achieves full channel diversity. Third, the channel representation in the DD domain is more sparse than that in the TF domain, such that a significantly smaller number of parameters are required to describe the channel, which leads to more simplified channel estimation. 
	
	Motivated by these advantages, many previous studies, e.g., \cite{raviteja2018interference,thaj2020low}, have investigated the performance of OTFS modulation in the time-varying channels. In \cite{raviteja2018interference}, the authors derived an input-output relation describing the OTFS modulation and demodulation, where both an ideal waveform and a rectangular waveform for pulse shaping are considered. The authors in \cite{thaj2020low} proposed an zero padding (ZP)-assisted OTFS system by placing guard null symbols to avoid the interference in the time domain, where a low complexity iterative decision feedback equalizer based on maximum ratio combining is proposed for symbol detection. Notably, \cite{raviteja2018interference,thaj2020low} only considered a single-input single-output wireless system. Some previous studies, e.g., \cite{khammammetti2018otfs,augustine2019interleaved}, have considered the application of OTFS over MAC. In \cite{khammammetti2018otfs}, the authors considered a uplink time-varying MAC, where OTFS is deployed at each device and the multi-user interference (MUI) is avoided by allocating the symbols at different DD resource blocks. The authors in \cite{augustine2019interleaved} allocated different TF resource blocks for each device to achieve a MUI-free reception at the RX, which effectively increases the error performance and reduces the detection complexity. 
	
	One notable advantage of applying OTFS into AirComp is that it can effectively transform designing AirComp over time-varying channels into time-invariant channels, since time-varying channels have almost constant channel gain in the DD domain. This advantage enables the utilization of consistent CSI for pre-processing, effectively overcoming the challenges posed by large signaling overhead and channel coherence times shorter than the signaling time. Therefore, motivated by these benefits, some recent studies, e.g., \cite{zhou2023otfs,zhou2023power}, have applied OTFS into the AirComp when the channel between each device and FC is multipath time-varying. The authors in \cite{zhou2023otfs} performed the channel estimation and optimized the precoding matrix based on the estimated channel to achieve the minimum MSE. Compared to \cite{zhou2023otfs}, the authors in \cite{zhou2023power} included a power control problem at each device. Although previous studies stand on their own merits, there are some limitations for \cite{zhou2023otfs,zhou2023power} in error performance and computational complexity. In particular, \cite{zhou2023otfs,zhou2023power} did not properly design the post-processing for a reliable target function estimation at the FC, and the computational complexity can be very high for a large OTFS frame size. 
	
	In this paper, we consider AirComp over multiple-access time-varying channels, where mobile devices non-orthogonally transmit their data over the same radio resources to the FC for averaging. For each device, the data is first modulated on the DD domain and then transmitted over the channel. In contrast to the OTFS-based multiple access schemes explored in \cite{khammammetti2018otfs, augustine2019interleaved}, where data allocation occurs at various DD resources, our approach involves utilizing the entire DD frame to modulate data at each device. This choice is motivated by the fact that AirComp utilizes the superposition property for the function estimation, which ultimately enhances communication efficiency. Moreover, we consider a multipath channel between each device and the FC, where different paths have different delay and Doppler shifts. After the signals are received by the FC, the aggregated signal is demodulated and estimated on the DD domain. Our objective is minimizing the function estimation error, which is measured by the MSE. We note that due to multipath channels, the MSE of OTFS-based AirComp is not only influenced by noise but also by inter-symbol interference (ISI) and inter-link interference (ILI). 
	
	In particular, we propose three schemes to estimate the target function at the FC. The first scheme (S1) directly estimates the target function under the influence of the noise and interference. In this scheme, we consider a power control problem, where we jointly optimize the transmit power at each device and a signal scaling factor, i.e., denoising factor, at FC, subject to individual power constraint at each device. In the second scheme (S2), to mitigate the impact of interference on the estimation, we propose using ZP-assisted OTFS for data modulation, where some null symbols are placed to avoid interference. In this scheme, we jointly optimize the transmit power, the denoising factor, and a weight for interference mitigation. In the third scheme (S3), by utilizing the full diversity property of OTFS, we estimate the  target function in a matrix form, even though the FC and all devices are equipped with a single antenna. In this scheme, we jointly optimize the precoding matrix at each device and the receiving filter matrix at the FC. 
	
	In summary, our contributions are summarized as follows:
	\begin{itemize}
		\item[1)] In S1, we derive closed-form expressions for the optimal transmit power and denosing factor that minimize the MSE. In S2, we propose a successive interference cancellation (SIC)-based algorithm for the target function estimation. In this algorithm, we derive the closed-form expressions for the transmit power, denosing factor, and the weight for interference cancellation. In S3, we rephrase the input-output relation of OTFS-based AirComp as a matrix form. We then propose an iterative algorithm that jointly optimize the precoding matrix and receiving filter matrix to achieve a minimum MSE.
		\item[2)] Numerical results validate the superiority of the optimal power allocation in S1. They also show that ZP-assisted OTFS in S2 can achieve a much lower MSE than S1, which validates the efficiency of the proposed SIC-based algorithm in mitigating the impact of interference. Moreover, this efficiency becomes more pronounced with a higher signal-to-noise ratio (SNR) and a fewer channel paths. Furthermore, numerical results validate the convergence of the proposed iterative algorithm in S3. 
		\item[3)] In the numerical results, we compare S1, S2, and S3 with benchmarks in \cite{zhou2023otfs,zhou2023power} from the perspectives of MSE and computational complexity. Numerical results show that S3 achieves significantly better error performance and a higher computational complexity than benchmarks, S1, and S2, where the computational complexity of S3 is largely determined by the OTFS frame size. Moreover, compared to benchmarks, S2 achieves a better error performance and a much lower computational complexity, and S1 achieves a better MSE for low SNR and a much lower computational complexity. Numerical results also show that the computational complexity of S2 is dominant by the number of channel paths, whereas S1 exhibits robustness for OTFS frame size and the number of channel paths.
	\end{itemize}
	
	This paper extends our preliminary work in \cite{huang2022} as follows. First, we include the scenario that at least two paths share the same delay in S2, while \cite{huang2022} considered each path has a different delay only. Second, we propose an iterative algorithm in S3 that jointly optimizing the precoding matrix and receiving filter matrix, while \cite{huang2022} did not consider this scheme. Third, we compare the performance of the proposed schemes with benchmarks in terms of MSE and computational complexity, while \cite{huang2022} did not perform this comparison.
	
	The reminder of this paper is organized as follows. In Section \ref{sm}, we describe the system model. In Section \ref{o}, we propose S1 and derive the closed-form expressions for the optimal transmit power and denoising factor. In Section \ref{i}, we propose S2 and the SIC-based algorithm. In Section \ref{ie}, propose S3 and the iterative algorithm. In Section \ref{n}, we present our numerical results. In Section \ref{c}, we draw some conclusions.
	
	\section{System Model}\label{sm}
	\begin{figure*}[!t]
		\begin{center}
			\includegraphics[width=2\columnwidth]{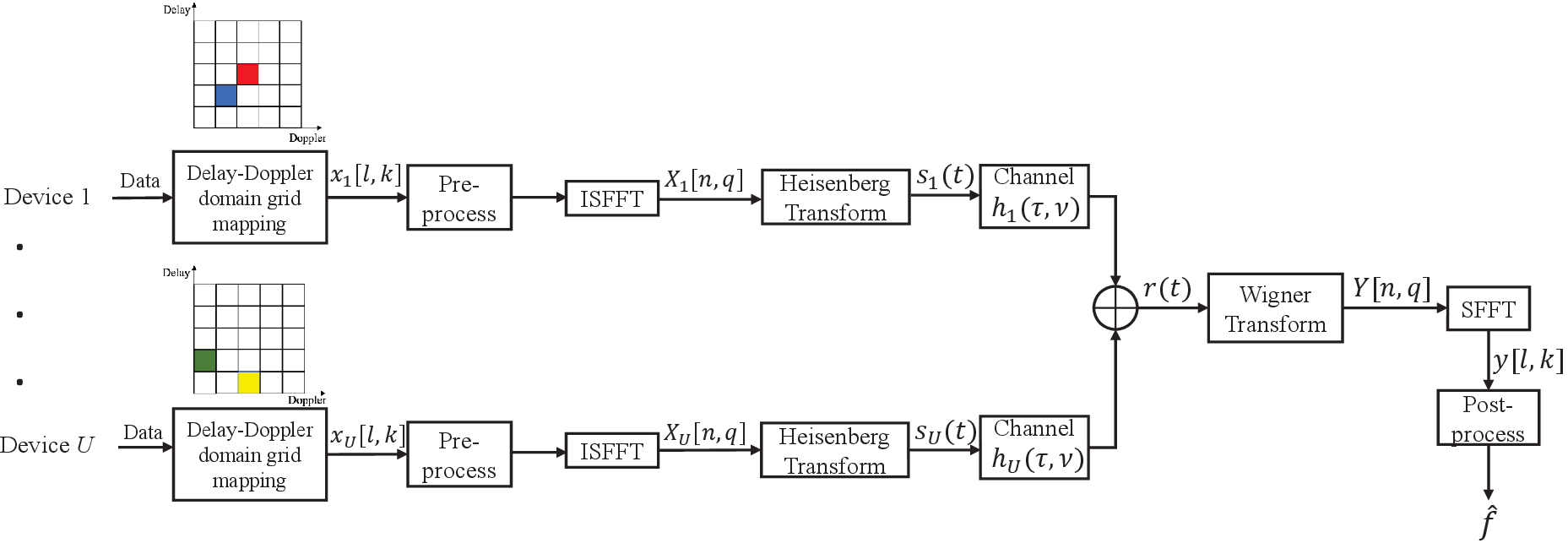}
			\caption{Illustration of the OTFS-based AirComp system.}\label{sys}\vspace{-0.5em}
		\end{center}\vspace{-4mm}
	\end{figure*}
	\subsection{OTFS Modulation \& Demodulation}
	We denote $x_u[l,k]\in\mathbb{C}$ as the transmitted data on the $l$-th row and $k$-th column of the two-dimensional (2D) DD grid for the $u$-th device, where $l\in\left\{0,1,..., M-1\right\}$ refers to delay, $k\in\left\{0,1,..., N-1\right\}$ refers to Doppler, $u$ is any device from the set $\mathcal{U}=\left\{1, ..., U\right\}$, $M$ is the number of subcarriers, and $N$ is the number of time slots. Then, the total number of transmitted data at each device per OTFS frame is $MN$. As in \cite{cao2020optimized}, we assume $\E\left[x_u[l,k]\right]=0$, $\E\left[x_u^2[l,k]\right]=1$, and $\E\left[x_u[l_1, k_1]x_u[l_2, k_2]\right]=0, \forall k_1\neq k_2, l_1\neq l_2$. A detailed OTFS-based AirComp system is shown in Fig. \ref{sys}, where each device and the FC are equipped with a single antenna. At the TX side, the $u$-th device first maps $x_u[l,k]$ to $X_u[n,q]$ on the TF domain using the inverse symplectic finite Fourier transform (ISFFT) as 
	\begin{align}
		X_u[n, q]=\frac{1}{\sqrt{NM}}\sum_{k=0}^{N-1}\sum_{l=0}^{M-1}x_u[l,k]e^{j2\pi\left(\frac{nk}{N}-\frac{ql}{M}\right)},
	\end{align}
	where $n=0, ..., N-1$ refers to time and $q=0, ..., M-1$ refers to frequency. Then, $X_u[n, q]$ is transformed into a continuous time waveform $s_u(t)$ by the Heisenberg transform \cite{mecklenbrauker1989tutorial} as
	\begin{align}
		s_u(t)=\sum_{n=0}^{N-1}\sum_{q=0}^{M-1}X_u[n,q]g_\mathrm{tx}(t-nT)e^{j2\pi q\Delta f(t-nT)},
	\end{align}
	where $g_\mathrm{tx}(t)$ is the pulse shaping filter at the TX, $T$ represents the duration of each time slot, and $\Delta f$ is the subcarrier spacing. To ensure TF symbols are orthogonal to each other, we have $T\Delta f=1$.
	
	The signal $s_u(t)$ is transmitted over a time-varying channel with complex baseband channel impulse response (CIR) $h_u(\tau, \nu)$, which is given by
	\begin{align}
		h_u(\tau,\nu)=\sum_{i=1}^{R_u}h_{u,i}\delta(\tau-\tau_{u,i})\delta(\nu-\nu_{u,i}),
	\end{align}
	where $R_{u}$ is the number of propagation paths from the $u$-th device to FC, $h_{u,i}$, $\tau_{u,i}$, and $\nu_{u,i}$ represent the path gain, delay and Doppler shifts of the $i$-th path from the $u$-th device to FC, respectively, and $\delta(\cdot)$ is the Dirac delta function. We further assume that the delay and Doppler shifts are the integer multiples of $1/(M\Delta f)$ and $1/(NT)$, respectively, where $NT$ and $M\Delta f$ are the total duration and bandwidth of one OTFS frame \cite{raviteja2018practical}. We clarify that the fractional Doppler shift can be addressed by adding virtual integer taps in the DD domain channel \cite{fish2013delay} or by applying TF domain windows \cite{wei2021transmitter}. In practice, the non-fractional case can be achieved by using sufficiently large $M$ and $N$ \cite{raviteja2018interference}. Accordingly, we have $\tau_{u,i}=l_{u,i}/(M\Delta f)$ and $\nu_{u,i}=k_{u,i}/(NT)$,
	where $l_{u,i}\in\mathcal{L}_u$ and $k_{u,i}\in\mathcal{K}_u$ are the indices of delay and Doppler. Here, $\mathcal{L}_u$ and $\mathcal{K}_u$ represent the sets of $l_{u,i}$ and $k_{u,i}$ among all paths between the $u$-th device and FC. Given the velocity of the $u$-th device as $v_u$, $k_{u, i}$ can be computed as
	\begin{align}
		k_{u,i}=\frac{v_uf_\mathrm{c}\cos(\rho_{u,i})N}{c\Delta f},
\end{align}
where $f_\mathrm{c}$ is the carrier frequency, $\rho_{u,i}$, is the angle between the wave propagation direction of the $i$-th path and the motion direction of the $u$-th device, and $c$ is the speed of light. Without loss of generality, we assume that $l_{u,1}\leq...\leq l_{u,i}\leq...\leq l_{u,R_u}$. Subsequently, the received signal at the FC is given by
	\begin{align}
		r(t)\!=\!\sum_{u=1}^{U}\!\int\!\!\!\int h_u(\tau, \nu)s_u(t-\tau)e^{j2\pi\nu(t-\tau)}d\tau d\nu+w(t),
	\end{align}
	where $w(t)$ is the additive white Gaussian noise (AWGN) at the FC with zero mean and variance $\sigma^2$.
	
	At the FC side, $r(t)$ first goes through a matched filter with output
	\begin{align}
		Y(f,t)=\int g_\mathrm{rx}^\dag(t'-t)r(t')e^{-j2\pi f(t'-t)}dt',
	\end{align}
	and $Y[n,q]$ is obtained by sampling $Y(t,f)$ at $t=nT$ and $f=q\Delta f$ as $Y[n,q]=Y(t,f)\big|_{t=nT, f=q\Delta f}$, where $g_\mathrm{rx}(t)$ is the receive pulse and the superscript $\dag$ represents complex conjugate. Next, $y[l,k]$ is obtained by taking the SFFT on the samples $Y[n,q]$ as
	\begin{align}
		y[l,k]=\frac{1}{\sqrt{NM}}\sum_{n=0}^{N-1}\sum_{q=0}^{M-1}Y[n,q]e^{-j2\pi\left(\frac{nk}{N}-\frac{ql}{M}\right)}.
	\end{align}
	
	In this paper, we consider a practical scenario where both $g_\mathrm{tx}(t)$ and $g_\mathrm{rx}(t)$ adopt rectangular forms, as in \cite{raviteja2018interference,thaj2020low}. Under this condition, the input-output relation between $x_u[l,k]$ and $y[l,k]$ is given by \cite[Eq. (30)]{raviteja2018practical}
	\begin{align}\label{y}
		y[l,k]=&\sum_{u=1}^{U}\sum_{i=1}^{R_u}h_{u,i}\alpha_{u,i}[l,k]\hat{x}_{u,i}+w[l,k],
	\end{align}
	where $\hat{x}_{u,i}=x_u\left[[l-l_{u,i}]_M, [k-k_{u,i}]_N\right]$ and
	\begin{align}\label{a}
		\alpha_{u,i}[l,k]=\left\{\begin{array}{lr}
			e^{-j2\pi\frac{k}{N}}z^{k_{u,i}([l-l_{u,i}]_M)},\;\mathrm{if}\;l\leq l_{u,i},\\
			z^{k_{u,i}([l-l_{u,i}]_M)},\;\mathrm{if}\;l\geq l_{u,i},		
		\end{array}
		\right.
	\end{align}
	with $z=e^{\frac{j2\pi}{MN}}$, $[\cdot]_N$ denotes mod-$N$ operation, and $w[l,k]$ represents the noise on the DD domain.
	
	\subsection{OTFS-based AirComp System}\label{ob}
	From \eqref{y}, we observe that $y[l,k]$ is a combination of $x_u[l,k]$ that has been shifted according to the delay and Doppler shifts specific to each channel path. Thus, for the corresponding $x_u[l,k]$ to be successfully aggregated at the FC, one approach is to designate one path as the principal path. The corresponding $x_u[l,k]$ is then arranged on the DD grid in accordance with the delay and Doppler shifts of the principal path for each device. Without any loss of generality, we consider $h_{u,1}, \forall u\in\mathcal{U}$, as the principal path gain. After pre-processing, we multiply $b_u[l,k]$ to each $x_u[l,k]$, where $b_u[l,k]$ is the transmit coefficient. Since $|\alpha_{u,i}[l,k]|=1$, we set $b_u[l,k]=\sqrt{p_u}h_{u,1}^\dag\alpha_{u,1}^\dag[l,k]/|h_{u,1}|$, where  $0\leq p_u\leq P_\mathrm{s}$ is the transmit power of the $u$-th device for each data and $P_\mathrm{s}$ is the average power budget. By replacing $x_u[l,k]$ in \eqref{y} with $b_u[l,k]x_u[l,k]$, we rephrase \eqref{y} as 
	\begin{align}\label{ykls}
		y[l,k]=&\sum_{u=1}^{U}\sqrt{p_u}|h_{u,1}|\hat{x}_{u,1}+\underbrace{\sum_{u=1}^{U}\sum_{i=2}^{R_u}\sqrt{p_u}h_{u,i}}_{\mathrm{ISI}+\mathrm{ILI}}\notag\\&\underbrace{\times\frac{\alpha_{u,i}[l,k]\alpha_{u,1}^\dag[l,k] h_{u,1}^\dag}{|h_{u,1}|} \hat{x}_{u,i}}+w[l,k].
	\end{align}
	Compared to the conventional input-output relation of AirComp \cite[Eq. (5)]{cao2020optimized}, \eqref{ykls} suffers from additional ISI and ILI, which are caused by the delay and Doppler spread of multiple paths. From \eqref{ykls}, we observe that only the time-invariant path gain, delay shift, and Doppler shift of the principal path are required for pre-processing at each device.
	
	In this paper, our aim is to estimate the average of the transmitted values from devices, i.e., $f[l,k]=\sum_{u=1}^{U}\hat{x}_{u,1}/U, \forall l\in\left\{1, ..., M-1\right\}, k\in\left\{1, ..., N-1\right\}$. After obtaining $y[l,k]$, the FC estimates $f[l,k]$ as $\hat{f}[l,k]=y[l,k]/(U\sqrt{\eta})$, where $\eta\geq0$ is the denoising factor. We further denote $\epsilon$ as the MSE between $\hat{f}[l,k]$ and $f[l,k]$, which is given by
	\begin{align}\label{ep}
		\epsilon=\frac{1}{U^2}\E\left[\left(\frac{y[l,k]}{\sqrt{\eta}}-\sum_{u=1}^{U}\hat{x}_{u,1}\right)^2\right],
	\end{align}
	where the expectation is taken over $x_u[l,k]$ and the noise.
	
	We note that calculating the average function is essential in many applications. For instance, in intelligent transportation systems, vehicles share information like speed, position, and environmental conditions to compute the average speed or density of vehicles on a particular road segment in real-time. This can improve traffic light control, dynamic tolling, and congestion avoidance by reducing the need for centralized data processing. Similarly, in federated learning, mobile devices, e.g., smartphones, vehicles, or wearables, train local machine learning models and aggregate updates at the server. The server computes the average of the model gradients to update the global model.
	
	\section{OTFS-Based AirComp Design}\label{o}
	In this section, we propose S1, where we jointly optimize the transmit power $p_u$ and denoising factor $\eta$ to minimize the MSE defined in \eqref{ep}. 
	\subsection{Problem Formulation}
	By substituting \eqref{ykls} into \eqref{ep}, we obtain 
	\begin{align}\label{mm}
		\epsilon\!=\sum_{u=1}^{U}\left(\frac{\sqrt{p_u}|h_{u,1}|}{\sqrt{\eta}}-1\right)^2+\sum_{u=1}^{U}\sum_{i=2}^{R_u}\frac{p_u|h_{u,i}|^2}{\eta}+\frac{\sigma^2}{\eta},
	\end{align}
	where we natually assume $\E[x_u[l,k]w[l,k]]=0$ and omit $1/U^2$ for simplicity. Our objective is to minimize \eqref{mm} by jointly optimizing $p_u$ and $\eta$. We then formulate the optimization problem, referred to as $\mathcal{P}1$, as follows
	\begin{align}\label{mp}
		\mathcal{P}1: &\min_{p_u\geq0, \eta\geq0} \sum_{u=1}^{U}\left(\frac{\sqrt{p_u}|h_{u,1}|}{\sqrt{\eta}}-1\right)^2+\sum_{u=1}^{U}\sum_{i=2}^{R_u}\frac{p_u|h_{u,i}|^2}{\eta}\notag\\&+\frac{\sigma^2}{\eta},\;\;
		\mathrm{s.t.}\; p_u\leq P_\mathrm{s}, \forall u\in\mathcal{U}.
	\end{align}
	\begin{remark}
		Filters with ideal waveforms, i.e., $g_\mathrm{tx}(t)$ and $g_\mathrm{rx}(t)$, which satisfy the bi-orthogonal property \cite[Eq. (4)]{hadani2017orthogonal} at the TX and RX, represent another widely considered scenario in OTFS modulation. According to \cite[Eq. (6)]{raviteja2018practical}, the input-output relation between $x_u[l,k]$ and $y[l,k]$ under this condition is given by 
		\begin{align}
			y[l,k]=\sum_{u=1}^{U}\sum_{i=1}^{R_u}h_{u,i}x_u[[l-l_{u,1}]_M,[k-k_{u,1}]_N],
		\end{align}
		which leads to the same MSE as in \eqref{mm}.
	\end{remark}
	\subsection{Optimization of $p_u$ and $\eta$}\label{oo}
	We first optimize $p_u$ for any given $\eta\geq0$. Accordingly, $\mathcal{P}1$ is simplified as
	\begin{align}\label{mpu}
		&\min_{p_u\geq0} \sum_{u=1}^{U}\left(\frac{\sqrt{p_u}|h_{u,1}|}{\sqrt{\eta}}-1\right)^2+\sum_{u=1}^{U}\sum_{i=2}^{R_u}\frac{p_u|h_{u,i}|^2}{\eta},\notag\\
		&\mathrm{s.t.}\; p_u\leq P_\mathrm{s}, \forall u\in\mathcal{U}.
	\end{align}
	By taking the derivative of the objective function in \eqref{mpu} with respect to $p_u$, we calculate the optimal $p_u$, denoted by $\hat{p}_u$, that minimizes \eqref{mpu} without accounting for the power constraint as $\hat{p}_u=|h_{u,1}|^2\eta/(\sum_{i=1}^{R_u}|h_{u,i}|^2)^2$. Upon incorporating the power constraint, the optimal transmit power, denoted by $p_u^*$, is given by
	\begin{align}\label{p}
		p_u^*=\min\left(P_\mathrm{s}, \frac{|h_{u,1}|^2\eta}{\left(\sum_{i=1}^{R_u}|h_{u,i}|^2\right)^2}\right).
	\end{align}
	
	We then optimize $\eta$. By substituting $p_u^*$ into \eqref{mp}, we obtain the following optimization problem, denoted by $\mathcal{P}2$, as 
	\begin{align}\label{m}
		\mathcal{P}2: &\min_{\eta\geq0}\sum_{u=1}^{U}\left(\min\left(\frac{\sqrt{P_\mathrm{s}}|h_{u,1}|}{\sqrt{\eta}}, \frac{|h_{u,1}|^2}{\sum_{i=1}^{R_u}|h_{u,i}|^2}\right)-1\right)^2\notag\\&+\sum_{u=1}^{U}\sum_{i=2}^{R_u}\min\left(\frac{P_\mathrm{s}|h_{u,i}|^2}{\eta}, \left(\frac{|h_{u,1}||h_{u,i}|}{\sum_{e=1}^{R_u}|h_{u,e}|^2}\right)^2\right)+\frac{\sigma^2}{\eta}.
	\end{align}
	Our next goal is to remove the $\mathrm{min}(\cdot)$ function in \eqref{m}. To this end, without loss of generality, we assume that
	\begin{align}
		\frac{\sum_{i=1}^{R_1}|h_{1,i}|^2}{|h_{1, 1}|}\leq...\leq\frac{\sum_{i=1}^{R_u}|h_{u,i}|^2}{|h_{u, 1}|}\leq...\leq\frac{\sum_{i=1}^{R_U}|h_{U,i}|^2}{|h_{U, 1}|}.
	\end{align}
	Based on this assumption, we divide the value of $\eta$ into $U+1$ intervals and define the $u$-th interval as
	\begin{align}\label{I}
		I_u\!\!=\!&\left\{\!\eta\Bigg|P_\mathrm{s}\!\left(\!\frac{\sum_{i=1}^{R_u}|h_{u,i}|^2}{|h_{u, 1}|}\!\right)^2\!\!\!\leq\eta\!\leq\! P_\mathrm{s}\!\left(\!\frac{\sum_{i=1}^{R_{u+1}}|h_{u+1,i}|^2}{|h_{u+1, 1}|}\right)^2\right\}, \notag\\&\forall u\in\left\{0, \mathcal{U}\right\},
	\end{align}
	where we define $\frac{\sum_{i=1}^{R_0}|h_{0,i}|^2}{|h_{0, 1}|}=0$ and $\frac{\sum_{i=1}^{R_{U+1}}|h_{U+1,i}|^2}{|h_{U+1, 1}|}=\infty$. Then, for $\eta\in I_u$, $\mathcal{P}2$ can be rephrased as 
	\begin{align}\label{me}
		&\min_{\eta\in I_u}H_u(\eta)= \sum_{j=1}^{u}\left(\frac{\sqrt{P_\mathrm{s}}|h_{j,1}|}{\sqrt{\eta}}-1\right)^2\notag\\&+\sum_{j=u+1}^{U}\left(\frac{|h_{j,1}|^2}{\sum_{i=1}^{R_j}|h_{j,i}|^2}-1\right)^2+\sum_{j=1}^{u}\sum_{i=2}^{R_j}\frac{P_\mathrm{s}|h_{j,i}|^2}{\eta}\notag\\&+\sum_{j=u+1}^{U}\sum_{i=2}^{R_j}\frac{|h_{j,1}|^2|h_{j,i}|^2}{\left(\sum_{e=1}^{R_u}|h_{u,e}|^2\right)^2}+\frac{\sigma^2}{\eta}.
	\end{align}
	We denote $\hat{\eta}_u>0$ as the optimal $\eta$ that achieves the minimum $H_u(\eta)$. In \eqref{me}, $H_u(\eta)$ first decreases on $[0, \hat{\eta}_u]$, and then increases on $[\hat{\eta}_u, \infty]$. By computing $\partial H_u(\eta)/\partial\eta=0$, we obtain $\hat{\eta}_u=((\sum_{j=1}^{u}\sum_{i=1}^{R_j}P_\mathrm{s}|h_{j,i}|^2+\sigma^2)/\sum_{j=1}^{u}\sqrt{P_\mathrm{s}}|h_{j,1}|)^2$. We then denote $\eta_u^*\in I_u$ as the optimal $\eta$ that minimizes \eqref{me}. According to \eqref{I} and $\hat{\eta}_u$, $\eta_u^*$ is given by
	\begin{align}\label{et}
		\eta_u^*=&\min\left[P_\mathrm{s}\left(\frac{\sum_{i=1}^{R_{u+1}}|h_{u+1,i}|^2}{|h_{u+1, 1}|}\right)^2\right.\notag\\&\left., \max\left(\hat{\eta}_u, P_\mathrm{s}\left(\frac{\sum_{i=1}^{R_u}|h_{u,i}|^2}{|h_{u, 1}|}\right)^2\right)\right].
	\end{align}
	We further denote $\eta^*$ as the optimal $\eta$ for $\mathcal{P}2$, which can be determined by comparing the values of $H_u(\eta_u^*)$ for $\eta_u^*$ within each interval. We suppose that $\eta^*=\eta^*_{u^*}$ and obtain $u^*$ as $u^*=\arg\min_{u\in\mathcal{U}}H_u(\eta_u^*)$. Then, based on \eqref{p},  \eqref{I}, and \eqref{et}, we derive and present $p_u^*$ and $\eta^*$ in the following theorem.
	\begin{theorem}\label{t1}
		The optimal $p_u$ and $\eta$ that jointly minimize $\mathcal{P}1$ are given by
		\begin{align}\label{pu}
			p_u^*=\left\{\begin{array}{lr}
				P_\mathrm{s}, \forall u\in\left\{1, ..., u^*\right\},\\
				\frac{|h_{u,1}|^2\eta^*}{\left(\sum_{i=1}^{R_u}|h_{u,i}|^2\right)^2}, \forall u\in\left\{u^*+1, ..., U\right\},
			\end{array}
			\right.
		\end{align}
		and
		\begin{align}\label{e}
			\eta^*=\left(\frac{\sum_{j=1}^{u^*}\sum_{i=1}^{R_j}P_\mathrm{s}|h_{j,i}|^2+\sigma^2}{\sum_{j=1}^{u^*}\sqrt{P_\mathrm{s}}|h_{j,1}|}\right)^2.
		\end{align}
	\end{theorem}
	\begin{IEEEproof}
		Please see Appendix \ref{A1}.
	\end{IEEEproof}

From Theorem \ref{t1}, it is evident that path gains are essential at both the FC and each device. In practical scenarios, the FC can estimate channel parameters based on the signals transmitted by the devices. This can be achieved by arranging pilot symbols on the DD grid, alongside guard and data symbols. The FC can then utilize methods such as the threshold approach \cite{raviteja2019embedded} or off-grid sparse Bayesian learning \cite{wei2022off} to estimate the channel information. Once the channel information is acquired, the FC transmits it back to the devices, enabling them to perform the pre-processing.
	\section{Interference Cancellation for ZP-Assisted OTFS-Based AirComp}\label{i}
	In this section, we propose S2, where we introduce the ZP-assisted OTFS and investigate its application in AirComp. We propose a novel algorithm for the ZP-assisted OTFS-based AirComp framework, which applies SIC for the target function estimation at the FC. In the algorithm, we optimize the transmit power, the denoising factor, and the weight for interference cancellation.
	\subsection{ZP-Assisted OTFS}
	We denote $\boldsymbol{X}_u\in\mathbb{C}^{M\times N}$ and $\boldsymbol{Y}\in\mathbb{C}^{M\times N}$ as the two-dimensional (2D) transmitted and received symbol matrices in the DD grid, respectively. In \cite{thaj2020low}, the authors proposed ZP-assisted OTFS, where the symbols in the last $l_{u, \mathrm{max}}$ rows of $\boldsymbol{X}_u$ are set to zero to avoid inter-block interference in the time domain. Here, $l_{u, \mathrm{max}}$ is the maximum channel delay spread index between the $u$-th device and FC. We then denote $\boldsymbol{x}_{u,m}\in\mathbb{C}^{N\times 1}$ and $\boldsymbol{y}_m\in\mathbb{C}^{N\times 1}$ as the column vectors that contain the symbols in the $m$-th row of $\boldsymbol{X}_u$ and $\boldsymbol{Y}$, respectively, i.e., $\boldsymbol{x}_{u,m}=\left[\boldsymbol{X}_u[m,0], \boldsymbol{X}_u[m,1], ..., \boldsymbol{X}_u[m, N-1]\right]^\mathrm{T}$ and $\boldsymbol{y}_m=\left[\boldsymbol{Y}[m,0], \boldsymbol{Y}[m,1], ..., \boldsymbol{Y}[m, N-1]\right]^\mathrm{T}$. We further denote $\tilde{\boldsymbol{x}}_u\in\mathbb{C}^{NM\times 1}$ and $\tilde{\boldsymbol{y}}\in\mathbb{C}^{NM\times 1}$ as the vectors of the transmitted symbols at the $u$-th device and the received symbols at FC, respectively, where $\tilde{\boldsymbol{y}}=\left[\boldsymbol{y}_0^\mathrm{T}, \boldsymbol{y}_1^\mathrm{T}, ..., \boldsymbol{y}_{M-1}^\mathrm{T}\right]^\mathrm{T}$ and $\tilde{\boldsymbol{x}}_u=\left[\boldsymbol{x}_{u,0}^\mathrm{T}, \boldsymbol{x}_{u,1}^\mathrm{T}, ..., \boldsymbol{x}_{u,M-1}^\mathrm{T}\right]^\mathrm{T}$. Accordingly, the relation between $\tilde{\boldsymbol{x}}_u$ and $\tilde{\boldsymbol{y}}$ is given by $\tilde{\boldsymbol{y}}=\sum_{u=1}^{U}\tilde{\boldsymbol{H}}_u\tilde{\boldsymbol{x}}_u+\tilde{\boldsymbol{w}}$, where $\tilde{\boldsymbol{H}}_u\in\mathbb{C}^{MN\times MN}$ is the channel matrix between the $u$-th device and FC in the DD domain and $\boldsymbol{w}$ is the AWGN vector with independent zero-mean elements of variance $\sigma^2$.
	\begin{figure}[!t]
		\begin{center}
			\includegraphics[width=1\columnwidth]{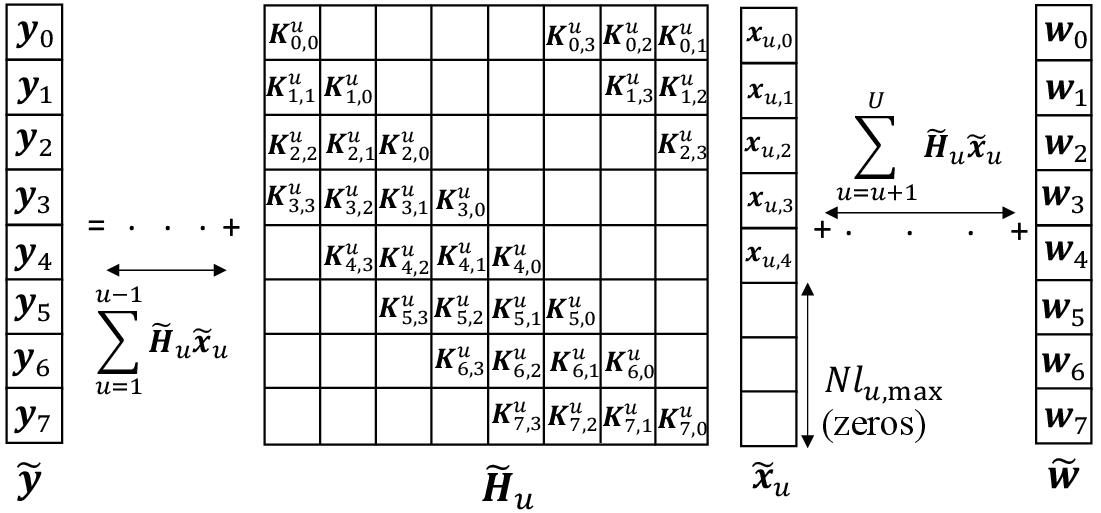}
			\caption{DD domain input-output relation for $N=M=8$, $R_u=4$, $l_u=[0,1,2,3]$, and $k_u=[0,1,2,3], \forall u\in\mathcal{U}$.}\label{mf}\vspace{-0.5em}
		\end{center}\vspace{-4mm}
	\end{figure}
	In Fig. \ref{mf}, we show a configuration example, where $\boldsymbol{K}^u_{m, l}\in\mathbb{C}^{N\times N}$ within $\tilde{\boldsymbol{H}}_u$ for $l\in\mathcal{L}_u$. We note that $\boldsymbol{K}_{m,l}^u$ occupies the rows of $\tilde{\boldsymbol{H}}_u$ from $mN+1$ to $(m+1)N$ and columns of $\tilde{\boldsymbol{H}}_u$ from is given by 
	\begin{align}\label{bk}
		&\boldsymbol{K}_{m,l}^u=\begin{bmatrix}
			\boldsymbol{\nu}_{m,l}^u(0) & \boldsymbol{\nu}_{m,l}^u(N-1) & \cdots & \boldsymbol{\nu}_{m,l}^u(1)\\
			\boldsymbol{\nu}_{m,l}^u(1) & \boldsymbol{\nu}_{m,l}^u(0) & \cdots & \boldsymbol{\nu}_{m,l}^u(2)\\
			\vdots & \vdots& \ddots & \vdots\\
			\boldsymbol{\nu}_{m,l}^u(N-1) & \boldsymbol{\nu}_{m,l}^u(N-2) & \cdots& \boldsymbol{\nu}_{m,l}^u(0)
		\end{bmatrix},
	\end{align}
	where
	\begin{align}\label{bn}
		\boldsymbol{\nu}_{m,l}^u(\kappa)\!=\!\left\{\begin{array}{lr}
			\!\!h_{u,i}z^{k(m-l)},\mathrm{if}\;l\!=\!l_{u,i}, k\!=\!k_{u,i}, \mathrm{and}\;\kappa=[k_{u,i}]_N,\\
			\!\!0,\mathrm{otherwise}.
		\end{array}
		\right.
	\end{align}
	In particular, $\boldsymbol{K}^u_{m, l}$ occupies $\tilde{\boldsymbol{H}}_u$ within the rows from $mN+1$ to $(m+1)N$ and columns from $([m+1-l]_M-1)N+1$ to $[m+1-l]_mN$. According to \eqref{bk} and Fig. \ref{mf}, it is evident that the configuration of $\tilde{\boldsymbol{H}}_u$ is influenced by Doppler shifts at each path. For different typologies of $\tilde{\boldsymbol{H}}_u$ at each device, we need to arrange the sequence of data accordingly to ensure that the corresponding $\boldsymbol{x}_u$ can be aggregated. In this section, we explore a scenario where all devices experience the same Doppler shift, allowing efficient SIC. This situation can occur when devices move together as a group, ensuring they travel in the same environment at the same speed. A practical example of this is multiple devices, such as smartphones, being on the same vehicle when transmitting data to the FC. Another relevant case is in satellite communications, where a satellite serves as the FC, receiving signals from ground-based devices. As satellites typically move at very high velocities, e.g., $7-8\;\mathrm{km}/\mathrm{s}$, it dominates the relative velocity between the satellite and the devices on the ground, given that the speed of ground vehicles is usually around $20-100\;\mathrm{m/s}$. Therefore, we can regard a same Doppler shift for all devices if they are moving in a similar direction relative to the satellite's motion, e.g., both moving towards or away from the satellite \cite{ali1998doppler}.
	\subsection{Algorithm for Interference Cancellation}\label{ef}
	In this section, we propose a new estimation algorithm for recovering the target function when applying ZP-assisted OTFS. As the last $Nl_{\mathrm{max}}$ elements in $\boldsymbol{x}_u$ are always set as zero as shown in Fig. \ref{mf}, $\sum_{u=1}^{U}\boldsymbol{x}_{u,0}$ and $\sum_{u=1}^{U}\boldsymbol{x}_{u, M-l_\mathrm{max}-1}$ can always be estimated without any interference. Moreover, as the estimation of $\sum_{u=1}^{U}\boldsymbol{x}_{u,2}$ in Fig. \ref{mf} experiences the most interference, its estimation is deferred to the last position. Therefore, a prioritized sequence for estimating $\sum_{u=1}^{U}\boldsymbol{x}_{u,m}, \forall m\in\left\{0, ..., M-l_\mathrm{max}-1\right\}$, is given by $\left[\boldsymbol{x}_{u,0}, \!\boldsymbol{x}_{u, 1},\! ...,\! \boldsymbol{x}_{u,m^*}, \!x_{u, M-l_\mathrm{max}-1}, \!x_{u, M-l_\mathrm{max}-2}, ..., \boldsymbol{x}_{u,m^*+1}\right]$. We note that $m^*$ is determined by delays among all paths.
	\begin{figure}[!t]
		\centering
		
		\subfigure[Different delay index in each path]{
			\begin{minipage}[t]{0.3\linewidth}
				\centering
				\includegraphics[width=0.95\columnwidth]{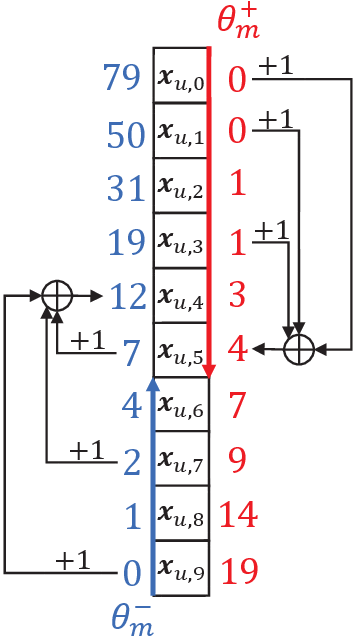}
				\label{p1}
			\end{minipage}%
		}%
		\subfigure[$l_{u,1}=l_{u,2}$]{
			\begin{minipage}[t]{0.3\linewidth}
				\centering
				\includegraphics[width=0.95\columnwidth]{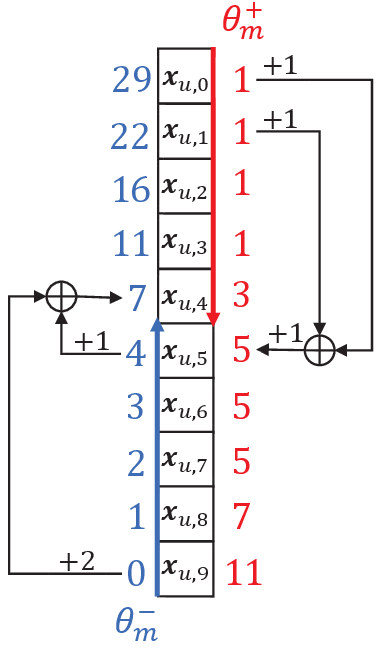}
				\label{p22}
			\end{minipage}%
		}
		\subfigure[$l_{u,2}=l_{u,3}$]{
			\begin{minipage}[t]{0.3\linewidth}
				\centering
				\includegraphics[width=0.95\columnwidth]{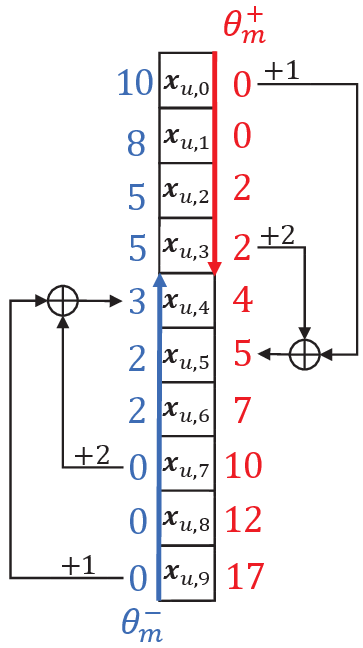}
				\label{p3}
			\end{minipage}%
		}
		\centering
		\caption{Illustration of the process for determining $\theta_m$ and $m^*$. (a): $\mathcal{L}_u=[0,2,4,5], m^*=5$. (b): $\mathcal{L}_u=[0,0,4,5]$, $m^*=4$. (c): $\mathcal{L}_u=[0,2,2,5]$, $m^*=3$.}
		\label{io}
	\end{figure}
	
	In Fig. \ref{io}, we show an example of how to determine $m^*$. We define an integer $\theta_m$ to measure the level of impact of the interference on the estimation of $\sum_{u=1}^{U}\boldsymbol{x}_{u,m}$, where a larger $\theta_m$ means more interference. For the estimation of $\sum_{u=1}^{U}\boldsymbol{x}_{u,m_1}$, the interference is either within the interval  $[\boldsymbol{x}_{u,0}, \boldsymbol{x}_{u,m_1-1}]$ or within $[\boldsymbol{x}_{u,m_1+1}, \boldsymbol{x}_{u,M-l_\mathrm{max}-1}]$. When the interference is within $[\boldsymbol{x}_{u,0}, \boldsymbol{x}_{u,m_1-1}]$, we set $\theta_{m_1}=\theta_{m_1}^+$. When the interference is within $[\boldsymbol{x}_{u,m_1+1}, \boldsymbol{x}_{u,M-l_\mathrm{max}-1}]$, we set $\theta_{m_1}=\theta_{m_1}^-$. As shown in Fig. \ref{io}, there are three scenarios for the delays among all paths. The first scenario is that each path has a different delay, as shown in Fig. \ref{p1}. For this scenario, we assume that there are $|\mathcal{M}^+|$ interference terms $\boldsymbol{x}_{u,m}$, $\forall m\in\mathcal{M}^+$, for the estimation of $\sum_{u=1}^{U}\boldsymbol{x}_{u,m_1}$, where $\mathcal{M}^+$ is a set containing the index $m$ of all interference terms within $[\boldsymbol{x}_{u,0}, \boldsymbol{x}_{u,m_1-1}]$. When the interference is within $[\boldsymbol{x}_{u,m_1+1}, \boldsymbol{x}_{u,M-l_\mathrm{max}-1}]$, we assume that there are $|\mathcal{M}^-|$ interference terms $\boldsymbol{x}_{u,m}$, $\forall m\in\mathcal{M}^-$. Then, we calculate $\theta_{m_1}^+$ and $\theta_{m_1}^-$ as $\theta_{m_1}^+=\sum_{m\in\mathcal{M}^+}\theta_m^++|\mathcal{M}^+|$ and $\theta_{m_1}^-=\sum_{m\in\mathcal{M}^-}\theta_m^-+|\mathcal{M}^-|$. When there is no interference for estimation, we set $\theta_{m_1}^+=\theta_{m_1}^-=0$. The second scenario is that the first path has the same delay with other paths, as shown in Fig. \ref{p22}. For this scenario, we assume that $r_\mathrm{s}$ paths have the same delay as the first path. Then, we calculate $\theta_{m_1}^+$ and $\theta_{m_1}^-$ as $\theta_{m_1}^+=\sum_{m\in\mathcal{M}^+}\theta_m^++|\mathcal{M}^+|+r_\mathrm{s}$ and $\theta_{m_1}^-=\sum_{m\in\mathcal{M}^-}\theta_m^-+|\mathcal{M}^-|+r_\mathrm{s}$. When there is no interference for estimation, we set $\theta_{m_1}^+=\theta_{m_1}^-=r_\mathrm{s}$. In the third scenario, $r_\mathrm{s}$ paths share a common delay, distinct from the delay in the first path. For this scenario, the calculation of $\theta_{m_1}^+$ and $\theta_{m_1}^-$ is the same as that in the first scenario. After obtaining $\theta_{m_1}^+$ and $\theta_{m_1}^-$, $m^*$ is determined as the maximum $m_1$ when $\theta_{m_1}^+\leq\theta_{m_1}^-$.  
	\subsubsection{Different $l_i$ associated to each path}\label{ee}
	In this subsection, we assume that $l_i$ on each path is different, i.e., $l_i\neq l_j, \forall i\neq j, 1\leq i,j\leq R$. Then, according to \eqref{bk} and \eqref{bn}, only one element in each row of $\boldsymbol{K}_{m,l}^u$ is non-zero. We first consider the scenario with \textit{no interference} during the estimation of $\sum_{u=1}^{U}\boldsymbol{x}_{u,m}$. To illustrate this, we use the estimation of $\sum_{u=1}^{U}\boldsymbol{x}_{u,0}$ from Fig. \ref{mf} as an example. We define $\gamma_m$ as any element within the set $\left\{mN, ..., (m+1)N-1\right\}$ and denote $y_{\gamma_m}$, $x_{u,\gamma_m}$, and $w_{\gamma_m}$ as the $\gamma_m$th element in $\boldsymbol{y}$, $\boldsymbol{x}_u$, and $\boldsymbol{w}$, respectively.
	As only $\boldsymbol{\nu}_{0,0}^u(0)=h_{u,1}$ in $\boldsymbol{K}_{0,0}^u$ is non-zero, the relationship between $y_{\gamma_0}$ and $x_{u, \gamma_0}$ is given by \begin{align}\label{yg}
		y_{\gamma_0}=\sum_{u=1}^{U}\sqrt{p_{u,0}}|h_{u,1}|x_{u,\gamma_0}+w_{\gamma_0},
	\end{align}
	where $\sqrt{p_{u, m}}$ is the transmit power for $x_{u,\gamma_m}$. We denote $\hat{f}_m$ as the estimated value of $\sum_{u=1}^{U}x_{u,\gamma_m}$ and $\epsilon_m$ as the MSE associated with estimating $\sum_{u=1}^{U}x_{u, \gamma_m}$. Here, $\hat{f}_0=y_{\gamma_0}/\sqrt{\eta_0}$, where $\eta_m$ is the denoising factor for $y_{\gamma_m}$. We note that \eqref{yg} has the same format as that of \cite[Eq. (5)]{cao2020optimized}, which suggests that the optimization procedures of $p_{u,0}$ and $\eta_0$ can follow \cite{cao2020optimized}. Thus, $p_{u,0}^*$ and $\eta_0^*$ are given by \cite[Eqs. (24), (25)]{cao2020optimized}
	\begin{align}\label{pug}
		p^*_{u, 0}=\left\{\begin{array}{lr}
			P_\mathrm{s},\;\forall u\in\left\{1, ..., u_0^*\right\},\\
			\frac{\eta^*_0}{|h_{u,1}|^2}, \forall u\in\left\{u_0^*+1, ..., U\right\},
		\end{array}
		\right.
	\end{align}
	and
	\begin{align}\label{eta}
		\eta^*_0=\left(\frac{\sigma^2+\sum_{j=1}^{u^*_0}P_\mathrm{s}|h_{j,1}|^2}{\sum_{j=1}^{u^*_0}\sqrt{P_\mathrm{s}}|h_{j,1}|}\right)^2,
	\end{align}
	respectively, with $u_0^*=\min_{u}\left(\frac{\sigma^2+\sum_{j=1}^{u}P_\mathrm{s}|h_{j,1}|^2}{\sum_{j=1}^{u}\sqrt{P}_\mathrm{s}|h_{j,1}|}\right)^2$. Here, the sequence of the devices is arranged based on $|h_{1,1}|^2\leq...\leq|h_{u,1}|^2\leq...\leq|h_{U,1}|^2$.
	
	Next, we consider the scenario where multiple interference terms impact the estimation of $\sum_{u=1}^{U}\boldsymbol{x}_{u,m}$. For illustrative purposes, we focus on the estimation of $\sum_{u=1}^{U}\boldsymbol{x}_{u,1}$ as an example. As depicted in Fig. \ref{mf}, $x_{u,\gamma_0}$ acts as an interference during the estimation of $\sum_{u=1}^{U}x_{u,\gamma_1}$ when examining the relation between $x_{u,\gamma_1}$ and $y_{\gamma_1}$. Given that we have estimated $\sum_{u=1}^{U}x_{u,\gamma_0}$, it can be subtracted from $y_{\gamma_1}$, leading to
	\begin{align}\label{yg1}
		y_{\gamma_1}=&\sum_{u=1}^{U}\sqrt{p_{u, 1}}|h_{u,1}|x_{u,\gamma_1}+\sum_{u=1}^{U}\frac{\sqrt{p_{u,0}^*}h_{u,2}h_{u,1}^\dag}{|h_{u,1}|}x_{u, \gamma_0}\notag\\&+w_{\gamma_1}-\zeta_{1,0}\hat{f}_0,
	\end{align}
	where $\zeta_{m,j}>0$ is a weight for $\hat{f}_j$ subtracted from the signal $y_{\gamma_m}$. By substituting \eqref{yg1} and $\hat{f}_0$ into \eqref{ep}, we formulate the optimization problem that minimizes the MSE between $\hat{f}_1$ and $\sum_{u=1}^{U}x_{u,\gamma_1}$ as
	\begin{align}\label{eps}
		\min_{p_{u,1}\geq0, \eta_1\geq0, \zeta_{1,0}\geq0}	&\epsilon_1=\!\sum_{u=1}^{U}\!\left(\frac{p_{u, 1}|h_{u,1}|}{\sqrt{\eta_1}}-1\right)^2+\frac{\E[G(\zeta_{1,0})^2]}{\eta_1},\notag\\&\mathrm{s.t.}\;p_{u,1}\leq P_\mathrm{s}, \forall u\in\mathcal{U},
	\end{align}
	where 
	\begin{align}
		G(\zeta_{1,0})=&\sum_{u=1}^{U}\left(\frac{h_{u,2}\sqrt{p_{u, 0}^*}h^\dag_{u,1}}{|h_{u,1}|}-\frac{\zeta_{1,0}\sqrt{p_{u, 0}^*}|h_{u,1}|}{\sqrt{\eta_0^*}}\right)x_{u, \gamma_0}\notag\\&+w_{\gamma_1}-\frac{\zeta_{1,0}w_{\gamma_0}}{\sqrt{\eta_0^*}}.
	\end{align}
	In the following theorem, we derive and present the optimal $p_{u,1}$, $\eta_1$, and $\zeta_{1,0}$ for \eqref{eps}.
	\begin{theorem}\label{t2}
		The optimal $p_{u, 1}$, $\eta_1$, and $\zeta_{1,0}$ that minimize $\epsilon_1$ are given by
		\begin{align}\label{pug1}
			p_{u,1}^*=\left\{\begin{array}{lr}
				P_\mathrm{s},\;\forall k\in\left\{1, ..., u_1^*\right\},\\
				\frac{\eta^*_1}{|h_{u,1}|^2}, \forall u\in\left\{u_1^*+1, ..., U\right\},
			\end{array}
			\right.
		\end{align}
		\begin{align}\label{egl}
			\eta_1^*=\left(\frac{\min\E[G(\zeta_{1,0})^2]+\sum_{j=1}^{u^*_1}P_\mathrm{s}|h_{j,1}|^2}{\sum_{j=1}^{u^*_1}\sqrt{P_\mathrm{s}}|h_{u,1}|}\right)^2,
		\end{align}
		and
		\begin{align}\label{d}
			\zeta_{1,0}^*=\frac{\sqrt{\eta_0^*}\sum_{u=1}^{U}|h_{u,2}||h_{u,1}|p_{u,0}^*}{\sum_{u=1}^{U}p_{u,0}^*|h_{u,1}|^2+\sigma^2},
		\end{align}
		where
		\begin{align}
			\min\E[G(\zeta_{1,0})^2]=&\sum_{u=1}^{U}|h_{u,2}|^2p_{u,0}^*+\sigma^2\notag\\&-\frac{(\sum_{u=1}^{U}|h_{u,2}||h_{u,1}|p_{u,0}^*)^2}{\sum_{u=1}^{U}p_{u,0}^*|h_{u,1}|^2+\sigma^2},
		\end{align}
		and
		$u_1^*=\min_u\left(\frac{\min\E\left[G(\zeta_{1,0})^2\right]+\sum_{j=1}^{u}P_\mathrm{s}|h_{j,1}|^2}{\sum_{j=1}^{u}\sqrt{P_\mathrm{s}}|h_{j,1}|}\right)^2.$
	\end{theorem}
	\begin{IEEEproof}
		Please see Appendix \ref{A2}.
	\end{IEEEproof}
	
	Notably, the estimation of $\sum_{u=1}^{U}x_{u,\gamma_1}$ in \eqref{yg1} is influenced only by a single interference $x_{u,\gamma_0}$. Furthermore, we consider the scenario that the detection of $\sum_{u=1}^{U}x_{u,\gamma_{m_1}}$ is affected by more than one interference, denoted by $x_{\gamma_m}, \forall m\in\mathcal{M}$. In the following, we assume that the interference is within $[\boldsymbol{x}_0, \boldsymbol{x}_{u,m_1-1}]$. Then, the relation between $x_{u,\gamma_{m_1}}$ and $y_{\gamma_{m_2}}$ is given by
	\begin{align}
		&y_{\gamma_{m_2}}=\sum_{u=1}^{U}\sqrt{p_{u, m_1}}|\nu_{m_2,l_1}^u|x_{u,\gamma_{m_1}}\notag\\&+\sum_{m\in\mathcal{M}}\sum_{u=1}^{U}\sqrt{p_{u,m}^*}\nu_{m_2, m_1-m+l}^u\frac{(\nu_{m_2,l_1}^u)^\dag x_{u,\gamma_{m}}}{|\nu_{m_2,l_1}^u|}+w_{\gamma_{m_2}}\notag\\&-\sum_{m\in\mathcal{M}}\zeta_{m_2, m}\hat{f}_{m},
	\end{align}
	where $\nu_{m,m_1-r+l}^u$ represents the non-zero element in $\boldsymbol{K}_{m,m_1-r+l}^u$ that corresponds to $x_{u,\gamma_m}$. We then formulate the optimization problem based on the MSE between $\hat{f}_{m_1}$ and $\sum_{u=1}^{U}x_{u,\gamma_{m_1}}$ as
	\begin{align}\label{mpur}
		&\min_{p_{u,m_1}\geq0, \eta_{m_2}\geq0,\zeta_{m_2, m}\geq0, \forall m\in\mathcal{M}}\!\epsilon_{m_1}\!\!=\!\!\!\sum_{u=1}^{U}\!\!\left(\!\!\frac{\sqrt{p_{u, m_1}}|\nu_{m_2, l_1}^u|}{\sqrt{\eta_{m_2}}}\!-\!1\!\!\right)^2\notag\\&+\frac{\sum_{m\in\mathcal{M}}\E\left[G(\zeta_{m_2, m})^2\right]}{\eta_{m_2}},\;\mathrm{s.t.}\;p_{u, m_1}\leq P, \forall u\in\mathcal{U},
	\end{align}
	where $E\left[G(\zeta_{m_2, m})^2\right]$ is a quadratic function relating to $\zeta_{m_2, m}$. Since \eqref{mpur} has a same format as that of \eqref{eps}, optimal values $p_{u,m_1}^*$, $\eta_{m_2}^*$, and $\zeta_{m_2, m}^*$ can be derived using the methodology outlined in Appendix \ref{A2}. We note that the optimization problem for the interference within $[\boldsymbol{x}_{u,m_1+1}, \boldsymbol{x}_{u,M-l_\mathrm{max}-1}]$ is similar to \eqref{mpur}, which we omit here due to the page limit.
	
	Based on Theorem \ref{t2}, it is evident that in addition to requiring the channel gains at both the FC and the devices to determine the optimal transmit power, denoising factor, and interference cancellation weights, the FC also needs the optimal transmit power used for transmitting the previous symbols. This necessitates that each device communicates its optimal transmit power to the FC. Furthermore, once the FC calculates the optimal denoising factor, it relays this information to all devices, enabling them to compute their respective optimal transmit powers.
	\subsubsection{At least two paths share the same $l$}
	In this subsection, we consider the scenario that at least two paths have the identical delay. As per \eqref{bk} and \eqref{bn}, several elements in each row of $\boldsymbol{K}_{m,l}^u$ are non-zero. We first consider the scenario that the first path shares the same delay with other paths. In this scenario, we first assess the condition that there is \textit{no interference} caused by different delays in other paths. We denote $\boldsymbol{\nu}_{m,l}^u(\varphi), \forall\varphi\in\mathcal{G}_l$, as all non-zero elements in each row of $\boldsymbol{K}_{m,l}^u$ and build the input-output relation between $x_{u,\gamma_{m_1}}$ and $y_{\gamma_{m_2}}$ as follows
	\begin{align}\label{ygm}
		y_{\gamma_{m_2}}\!\!=&\!\!\sum_{u=1}^{U}\!\sqrt{p_{u,m_1}}|\boldsymbol{\nu}_{m_2,l_1}^u(\varphi_1)|x_{u,\gamma_{m_1}}\!\!\!+\!\!\sum_{u=1}^{U}\sum_{\varphi\in\mathcal{G}_{l_1}\setminus\{\varphi_1\}}\!\!\!\sqrt{p_{u,m_1}}\notag\\&\times\frac{\boldsymbol{\nu}_{m_2,l_1}^u(\varphi)\left(\boldsymbol{\nu}_{m_2,l_1}^u(\varphi_1)\right)^\dag x_{u,\gamma_{m_1}}^\varphi}{|\boldsymbol{\nu}_{m_2,l_1}^u(\varphi_1)|}+w_{\gamma_{m_2}},
	\end{align}
	where $\boldsymbol{\nu}_{m_2,l_1}^u(\varphi_1)$ represents the item of $\boldsymbol{\nu}_{m_2,l_1}^u(\varphi)$ that multiplies with the data $x_{u,\gamma_{m_1}}$ that we want to estimate, $x_{u,\gamma_{m_1}}^\varphi$ represents the interference in $\boldsymbol{x}_{u,m_1}$, which is induced by the paths having the same delay as the first path, and $\mathcal{G}_{l_1}\setminus\{\varphi_1\}$ represents removing $\varphi_1$ from set $\mathcal{G}_{l_1}$. From \eqref{ygm}, we observe that the interference $x_{u,\gamma_{m_1}}^\varphi$ cannot be canceled since the value has not been estimated. We then derive the MSE for estimating $\sum_{u=1}^{U}x_{u,\gamma_{m_1}}$ and formulate the following optimization problem
	\begin{align}\label{er}
		\min_{p_{u,m_1}\geq0, \eta_{m_2}\geq0} &\epsilon_{m_1}=\sum_{u=1}^{U}\left(\frac{\sqrt{p_{u, m_1}}|\boldsymbol{\nu}_{m_2,l_1}^u(\varphi_1)|}{\sqrt{\eta_{m_2}}}-1\right)^2\notag\\&+\sum_{u=1}^{U}\sum_{\varphi\in\mathcal{G}_{l_1}\setminus\{\varphi_1\}}\frac{p_{u,m_1}|\boldsymbol{\nu}_{m_2,l}^u(\varphi)|^2}{\eta_{m_2}}+\frac{\sigma^2}{\eta_{m_2}}, \notag\\&\mathrm{s.t.}\;p_{u, m_1}\leq P_\mathrm{s}, \forall u\in\mathcal{U}. 
	\end{align}
	We note that \eqref{er} has a same format as \eqref{mp}. Thus, the optimization procedures of $p_{u,m_1}$ and $\eta_{m_2}$ follow Section \ref{oo}.
	
	Next, we consider the condition that there are \textit{multiple interferences} induced by delays different from the delay in the first path. We denote $\hat{f}_{m,\varphi}$ as the estimated value of $\sum_{u=1}^{U}x_{u,m}^\varphi$ and $\zeta_{m_2,m}^\varphi$ as the weight for subtracting $\hat{f}_{m,\varphi}$ from the received signal. Then, the relation between $x_{\gamma_{m_1}}$ and $y_{\gamma_{m_2}}$ is given by \eqref{byr}. We then formulate the optimization problem in \eqref{er2} by deriving the MSE between $\hat{f}_{m,\varphi}$ and $x_{\gamma_{m_1}}$. We note that \eqref{byr} and \eqref{er2} are presented on the top of the next page.
	\begin{figure*}
		\begin{align}\label{byr}
			y_{\gamma_{m_2}}=&\sum_{u=1}^{U}\sqrt{p_{u,m_1}}|\boldsymbol{\nu}_{m_2,l_1}^u(\varphi_1)|x_{\gamma_{m_1}}+\sum_{u=1}^{U}\sum_{\varphi\in\mathcal{G}_{l_1}\setminus\{\varphi_1\}}\frac{\sqrt{p_{u,m_1}}\boldsymbol{\nu}_{m_2,l_1}^u(\varphi)\left(\boldsymbol{\nu}_{m_2,l_1}^u(\varphi_1)\right)^\dag}{|\boldsymbol{\nu}_{m_2,l_1}^u(\varphi_1)|}x_{u,m_1}+\sum_{u=1}^{U}\sum_{m\in\mathcal{M}}\sum_{\varphi\in\mathcal{G}_{m_1-m+l_1}}\notag\\&\frac{\sqrt{p_{u,m}^*}\boldsymbol{\nu}_{m_2, m_1-m+l_1}^u(\varphi)\left(\boldsymbol{\nu}_{m_2-m_1+m, l_1}^u(\varphi_1)\right)^\dag}{|\boldsymbol{\nu}_{m_2-m_1+m, l_1}^u(\varphi_1)|}x_{u,m}^\varphi+w_{\gamma_{m_2}}-\sum_{m\in\mathcal{M}}\sum_{\varphi\in\mathcal{G}_{m_1-m+l_1}}\zeta_{m_2, m}^\varphi\hat{f}_{m,\varphi}.
		\end{align}
		\begin{align}\label{er2}
			&\min_{p_{u,m_1}\geq0, \eta_{m_2}\geq0, \zeta_{m_2, m}^\varphi\geq0, \forall m\in\mathcal{M}, \forall\varphi\in\mathcal{G}_{m_1-m+l_1}}
			\epsilon_{m_1}=\sum_{u=1}^{U}\left(\frac{\sqrt{p_{u, m_1}}|\boldsymbol{\nu}_{m_2,l_1}^u(\varphi_1)|}{\sqrt{\eta_{m_2}}}-1\right)^2+\sum_{u=1}^{U}\sum_{\varphi\in\mathcal{G}_{l_1}\setminus\{\varphi_1\}}\frac{p_{u,m_1}|\boldsymbol{\nu}_{m_2,l_1}^u(\varphi)|^2}{\eta_{m_2}}\notag\\&+\frac{\sum_{m\in\mathcal{M}}\sum_{\varphi\in\mathcal{G}_{m_1-m+l_1}}\E\left[G(\zeta_{m_2, m}^\varphi)^2\right]}{\eta_{m_2}}, \;\mathrm{s.t.}\;p_{u, m_1}\leq P_\mathrm{s}, \forall u\in\mathcal{U},
		\end{align}
		\hrulefill \vspace*{-1pt}
	\end{figure*}
	In \eqref{er2}, $\E\left[G(\zeta_{m_2, m}^\varphi)^2\right]$ is the quadratic function related to $\zeta_{m_2, m}^\varphi$. Therefore, the optimization procedure of $\zeta_{m_2, m}^\varphi$ follows Appendix \ref{A2}. After obtaining $\sum_{m\in\mathcal{M}}\sum_{\varphi\in\mathcal{G}_{m_1-m+l_1}}\E\left[G(\zeta_{m_2, m}^\varphi)^2\right]$, the expression format of \eqref{er2} is the same as that in \eqref{mp}. Therefore, the optimization procedures of $p_{u,m_1}$ and $\eta_{m_2}$ follow Section \ref{oo}.
	
	The second scenario is that the last path shares the same delay with other paths, and the analysis for this scenario is the same as that in the first scenario. Therefore, we omit the details here due to the page limit. The third scenario is that the middle paths share the same delay. For this scenario, all the interferences have been estimated before the target function estimation, such that the estimation procedures follow Section \ref{ee}. We omit the details here due to the page limit.
	\section{Iterative Estimation for OTFS-Based AirComp}\label{ie}
	In this section, we propose S3, where we consider a matrix form-based input-output relation for OTFS-based AirComp. At each device, we apply a precoding matrix for pre-processing, and a receiving filter matrix is applied at the FC for estimating the target function. We then propose an iterative algorithm to jointly optimize the precoding matrix and receiving filter matrix to achieve the minimum MSE.
	\subsection{Problem Formulation}
	We denote $\boldsymbol{x}_u\in\mathbb{C}^{MN\times 1}$ and $\boldsymbol{y}\in\mathbb{C}^{MN\times 1}$ as the column-wise vectorization of $\boldsymbol{X}_u$ and $\boldsymbol{Y}$. Then, \eqref{y} can be rephrased in a matrix format as $\boldsymbol{y}=\sum_{u=1}^{U}\boldsymbol{H}_u\boldsymbol{x}_u+\boldsymbol{w}$ \cite[Eq. (11)]{raviteja2018practical}, where $\boldsymbol{H}_u=(\boldsymbol{F}_N\otimes\boldsymbol{I}_M)\hat{\boldsymbol{H}}_u(\boldsymbol{F}_N^H\otimes\boldsymbol{I}_M)$ with $\boldsymbol{F}_N\in\mathbb{C}^{N\times N}$ represents an $N$-point FFT, $\boldsymbol{F}_N^H$ is an $N$-point IFFT, $\tilde{\boldsymbol{w}}\in\mathbb{C}^{MN\times 1}$ is the noise vector, and $\hat{\boldsymbol{H}}_u$ is given by $\hat{\boldsymbol{H}}_u=\sum_{i=1}^{R_u}h_{u,i}\boldsymbol{\Pi}^{l_{u,i}}\boldsymbol{\Delta}^{k_{u,i}}$, where $\boldsymbol{\Pi}\in\mathbb{C}^{MN\times MN}$ is the permutation matrix given by
	\begin{align}
		\boldsymbol{\Pi}=\begin{bmatrix}
			0 & \cdots & 0 & 1\\
			1 & \ddots & 0 & 0\\
			\vdots & \ddots& \ddots & \vdots\\
			0 & \cdots & 1 & 0
		\end{bmatrix},
	\end{align}
	and $\boldsymbol{\Delta}\in\mathbb{C}^{MN\times MN}$ is a diagonal matrix given by $\boldsymbol{\Delta}=\mathrm{diag}[z^0, z^1, ..., z^{MN-1}]$.
	
	In this section, our objective is to estimate $\sum_{u=1}^{U}\boldsymbol{x}_u/U$. To this end, we apply a linear pre-processing technique by multiplying $\boldsymbol{x}_u$ with a precoding matrix $\boldsymbol{B}_u$. Then, the input-output relation between $\boldsymbol{x}_u$ and $\boldsymbol{y}$ is given by
	\begin{align}\label{by}
		\boldsymbol{y}=\sum_{u=1}^{U}\boldsymbol{H}_u\boldsymbol{B}_u\boldsymbol{x}_u+\boldsymbol{w}.
	\end{align}
	As we do not need to select the principle path, we assume that the corresponding $x_u[l,k]$ is arranged at the same position in $\boldsymbol{X}_u$. At the FC side, we apply a linear equalizer to recover $\sum_{u=1}^{U}\boldsymbol{x}_u/U$. We denote $\hat{\boldsymbol{f}}$ as the estimated value of $\sum_{u=1}^{U}\boldsymbol{x}_u/U$ and we have $\hat{\boldsymbol{f}}=\boldsymbol{V}\boldsymbol{y}/U$, where $\boldsymbol{V}\in\mathbb{C}^{MN\times MN}$ is the receiving filter matrix at FC. Then, the MSE between $\hat{\boldsymbol{f}}$ and $\sum_{u=1}^{U}\boldsymbol{x}_u/U$ is given by
	\begin{align}\label{em}
		\epsilon=\frac{1}{U^2}\E\left[\Bigg|\Bigg|\boldsymbol{V}\boldsymbol{y}-\sum_{u=1}^{U}\boldsymbol{x}_u\Bigg|\Bigg|^2\right],
	\end{align}
	where $||\cdot||$ represents the Frobenius norm of an input variable.
	By substituting \eqref{by} into \eqref{em}, we obtain
	\begin{align}\label{ems}
		\epsilon=&\frac{1}{U^2}\left[\sum_{u=1}^{U}\mathrm{Tr}\left[\left(\boldsymbol{V}\boldsymbol{H}_u\boldsymbol{B}_u-\boldsymbol{I}\right)\left(\boldsymbol{V}\boldsymbol{H}_u\boldsymbol{B}_u-\boldsymbol{I}\right)^H\right]\right.\notag\\&\left.+\sigma^2\mathrm{Tr}\left(\boldsymbol{V}\boldsymbol{V}^H\right)\right],
	\end{align}
	where $\mathrm{Tr}(\cdot)$ represents trace and $\boldsymbol{V}^H$ represents the Hermitian transpose of $\boldsymbol{V}$. Our objective is to minimize \eqref{ems} by jointly optimizing $\boldsymbol{V}$ and $\boldsymbol{B}_u$. We assume that $P_\mathrm{t}$ is the total average power budget for transmitting $MN$ data at each device, and we have $P_\mathrm{t}=MNP_\mathrm{s}$. Then, we formulate the optimization problem, referred to as $\mathcal{P}3$, as follows
	\begin{align}\label{p2}
		&\mathcal{P}3: \min_{\boldsymbol{B}_u, \boldsymbol{V}} \sum_{u=1}^{U}\mathrm{Tr}\left[\left(\boldsymbol{V}\boldsymbol{H}_u\boldsymbol{B}_u-\boldsymbol{I}\right)\left(\boldsymbol{V}\boldsymbol{H}_u\boldsymbol{B}_u-\boldsymbol{I}\right)^H\right]\notag\\&+\sigma^2\mathrm{Tr\left(\boldsymbol{V}\boldsymbol{V}^H\right)}, \;\mathrm{s.t.}\; \mathrm{Tr}\left(\boldsymbol{B}\boldsymbol{B}^H\right)\leq P_\mathrm{t}, \forall u\in\mathcal{U}.
	\end{align}
	\subsection{Optimization of $\boldsymbol{B}_u$ and $\boldsymbol{V}$}
	Given the optimization function and power constraint in \eqref{p2}, we first construct a Lagrange dual objective function \cite{boyd2004convex} as follows
	\begin{align}
		&L(\boldsymbol{B}_u;\boldsymbol{V};\lambda_u)=\sum_{u=1}^{U}\mathrm{Tr}\left[\left(\boldsymbol{V}\boldsymbol{H}_u\boldsymbol{B}_u-\boldsymbol{I}\right)\left(\boldsymbol{V}\boldsymbol{H}_u\boldsymbol{B}_u-\boldsymbol{I}\right)^H\right]\notag\\&+\sigma^2\mathrm{Tr\left(\boldsymbol{V}\boldsymbol{V}^H\right)}+\sum_{u=1}^{U}\lambda_u\left(\mathrm{Tr}\left(\boldsymbol{B}_u\boldsymbol{B}_u^H\right)-P_\mathrm{t}\right),
	\end{align}
	where $\lambda_u$ is the Lagrange multiplier associated with the power constraint of device $u$. We further denote $\boldsymbol{B}_u^*$ and $\boldsymbol{V}^*$ as the primal optimal points and $\lambda_u^*$ as the dual optimal point with zero duality gap. Then, according to Karush-Kuhn-Tucker (KKT) conditions \cite{boyd2004convex}, we have $\lambda_u^*\geq0$, $\mathrm{Tr}\left(\boldsymbol{B}_u^*(\boldsymbol{B}_u^*)^H\right)-P_\mathrm{t}\leq0$,
	\begin{align}\label{lu}
		\lambda_u^*\left(\mathrm{Tr}\left(\boldsymbol{B}_u^*(\boldsymbol{B}_u^*)^H\right)-P_\mathrm{t}\right)=0,
	\end{align}
	and
	\begin{align}\label{f}
		\frac{\partial L(\boldsymbol{B}_u^*;\boldsymbol{V}^*;\lambda_u^*)}{\partial\boldsymbol{B}_u^*}=0,\;\frac{\partial L(\boldsymbol{B}_u^*;\boldsymbol{V}^*;\lambda_u^*)}{\partial\boldsymbol{V}^*}=0, \forall u\in\mathcal{U},
	\end{align}
	where we have
	\begin{align}\label{fp}
		\frac{\partial L(\boldsymbol{B}_u^*;\boldsymbol{V}^*;\lambda_u^*)}{\partial\boldsymbol{B}_u^*}=&\boldsymbol{H}_u^H(\boldsymbol{V}^*)^H\boldsymbol{V}^*\boldsymbol{H}_u\boldsymbol{B}_u^*-\boldsymbol{H}_u^H(\boldsymbol{V}^*)^H\notag\\&+\lambda_u^*\boldsymbol{B}_u^*,
	\end{align}
	and
	\begin{align}\label{fpl}
		\frac{\partial L(\boldsymbol{B}_u^*;\boldsymbol{V}^*;\lambda_u^*)}{\partial\boldsymbol{B}_u^*}=&\sum_{u=1}^{U}\left(\boldsymbol{V}^*\boldsymbol{H}_u\boldsymbol{B}_u^*(\boldsymbol{B}_u^*)^H\boldsymbol{H}_u^H\right.\notag\\&\left.-(\boldsymbol{V}_u^*)^H\boldsymbol{H}_u^H\right)+2\sigma^2\boldsymbol{V}^*.
	\end{align}
	
	According to \eqref{f}, \eqref{fp}, and \eqref{fpl}, we can derive $\boldsymbol{B}_u^*$ and $\boldsymbol{V}^*$ as
	\begin{align}\label{bu}
		\boldsymbol{B}_u^*=\left(\boldsymbol{H}_u^H(\boldsymbol{V}^*)^H\boldsymbol{V}^*\boldsymbol{H}_u+\lambda_u^*\boldsymbol{I}\right)^{-1}\boldsymbol{H}_u^H(\boldsymbol{V}^*)^H,
	\end{align}
	and
	\begin{align}\label{v}
		\boldsymbol{V}^*\!\!=\!\!\left(\!\sum_{u=1}^{U}\!(\boldsymbol{B}_u^*)^H\!\boldsymbol{H}_u^H\!\right)\!\!\!\left(\!\sum_{u=1}^{U}\!\!\boldsymbol{H}_u\!\boldsymbol{B}_u^*(\!\boldsymbol{B}_u^*\!)^H\!\boldsymbol{H}_u^H\!\!+\!\!\sigma^2\boldsymbol{I}\!\right)^{-1}.
	\end{align}
	\SetKwComment{Comment}{/* }{ */}
	\RestyleAlgo{ruled}
	\begin{algorithm}[!t]
		\caption{Iterative Algorithm for Optimizing $\boldsymbol{B}_u$ and $\boldsymbol{V}$}\label{alg2}
		Input: Channel matrix $\boldsymbol{H}_u$\;
		Initialize $\boldsymbol{B}_u(0)$ as the right singular matrix of $\boldsymbol{H}_u$\;
		Compute $\boldsymbol{V}(0)$ based \eqref{v} by replacing $\boldsymbol{B}_u^*$ with $\boldsymbol{B}_u(0)$\;
		\For{$\beta\gets1, 2, ..., \mathcal{B}$}{Compute $\lambda_u$ based on \eqref{s}\;
			\If{$\lambda_u<0$}{Set $\lambda_u=0$\;}Compute $\boldsymbol{B}_u(\beta)$, based on \eqref{bb} by replacing $\boldsymbol{V}^*$ with $\boldsymbol{V}(\beta-1)$\;Compute $\boldsymbol{V}(\beta)$ based on \eqref{v} by replacing $\boldsymbol{B}_u^*$ with $\boldsymbol{B}_u(\beta)$\;}Output: $\boldsymbol{B}_u^*$, $\boldsymbol{V}^*$
	\end{algorithm}
	From \eqref{bu} and \eqref{v}, we observe that $\boldsymbol{B}_u^*$ is determined by $\boldsymbol{V}^*$ and $\boldsymbol{V}^*$ is determined by $\boldsymbol{B}_u^*$. Therefore, we need to jointly optimize $\boldsymbol{B}_u$ and $\boldsymbol{V}$. An algorithm that iteratively updates $\boldsymbol{B}_u$ and $\boldsymbol{V}$ is proposed in Algorithm \ref{alg2}. We assume that there are $\mathcal{B}$ iterations required to obtain $\boldsymbol{B}_u^*$ and $\boldsymbol{V}^*$ and denote $\boldsymbol{B}_u(\beta)$ and $\boldsymbol{V}(\beta)$ as the value of $\boldsymbol{B}_u$ and $\boldsymbol{V}$ in the $\beta$-th round of iteration. Before the iteration, we initialize $\boldsymbol{B}_u(0)$ as the right singular matrix of $\boldsymbol{H}_u$ by performing singular value decomposition (SVD), and then normalize $\boldsymbol{B}_u(0)$ to make sure that $\mathrm{Tr}(\boldsymbol{B}_u(0)\boldsymbol{B}_u^H(0))\leq P_\mathrm{t}$. Moreover, in the $\beta$-th iteration, we define $\boldsymbol{T}_u(\beta)=\boldsymbol{H}_u^H\boldsymbol{V}^H(\beta)\boldsymbol{V}(\beta)\boldsymbol{H}_u$. As $\boldsymbol{T}_u(\beta)$ is a symmetric matrix, the SVD of $\boldsymbol{T}_u(\beta)$ is given by $\boldsymbol{T}_u(\beta)=\boldsymbol{\Lambda}_u(\beta)\boldsymbol{\Sigma}_u(\beta)\boldsymbol{\Lambda}_u^H(\beta)$ \cite{strang1993introduction}. Then, we can express $\mathrm{Tr}\left(\boldsymbol{B}_u(\beta)\boldsymbol{B}_u^H(\beta)\right)=\mathrm{Tr}\left[\left(\boldsymbol{\Sigma}_u(\beta)+\lambda_u\boldsymbol{I}\right)^{-2}\boldsymbol{\Gamma}_u(\beta)\right]$, where $\boldsymbol{\Gamma}_u(\beta)=\boldsymbol{\Lambda}_u^H(\beta)\boldsymbol{H}_u^H\boldsymbol{V}^H(\beta)\boldsymbol{V}(\beta)\boldsymbol{H}_u\boldsymbol{\Lambda}_u(\beta)$ and $(a)$ is achieved by substituting $\boldsymbol{T}_u(\beta)=\boldsymbol{\Lambda}_u(\beta)\boldsymbol{\Sigma}_u(\beta)\boldsymbol{\Lambda}_u^H(\beta)$ into the expression. According to \eqref{lu}, we can solve $\lambda_u$ numerically based on 
	\begin{align}\label{s}
		\sum_{i=1}^{MN}\frac{\Gamma_u^{[ii]}(\beta)}{\lambda_u+\Sigma_u^{[ii]}(\beta)}=P_\mathrm{t},
	\end{align}
	where $\Gamma_u^{[ii]}(\beta)$ and $\Sigma_u^{[ii]}$ represent the diagonal values of $\boldsymbol{\Gamma}_u(\beta)$ and $\boldsymbol{\Sigma}_u(\beta)$. If the derived $\lambda_u$ is less than 0, we set $\lambda_u=0$.
	
	As we already performed SVD for $\boldsymbol{T}_u(\beta)$, we can further express \eqref{bu} as
	\begin{align}\label{bb}
		\boldsymbol{B}_u^*=\boldsymbol{\Lambda}_u\left(\boldsymbol{\Sigma}_u+\lambda_u^*\boldsymbol{I}\right)^{-1}\boldsymbol{\Lambda}_u^H\boldsymbol{H}_u^H(\boldsymbol{V}^*)^H,
	\end{align}
	where the inverse of $\boldsymbol{\Sigma}_u+\lambda_u^*\boldsymbol{I}$ can be directly obtained by taking the reciprocal of the diagonal element. Additionally, the computational complexity of each iteration in Algorithm \ref{alg2} is $\mathcal{O}(U(M^3N^3+M^2N^2))$.
	
	\section{Numerical Results}\label{n}
	In this section, we evaluate the MSE performance and computational complexity of S1, S2, and S3. For S1, the MSE is computed by applying the optimal transmit power and the optimal denoising factor derived in Theorem \ref{t1}. For S2, the MSE is computed based on the SIC-based algorithm, where the optimal transmit power, denoising factor, and the weight for interference cancellation are derived in Theorem \ref{t2}. For S3, the MSE is computed based on the optimized precoding matrix and receiving filter matrix in Algorithm \ref{alg2}. Moreover, the evaluation focuses on the average MSE over a large number of channel realizations, which we set 1000 in the simulation. The path gains are randomly generated based on a uniform power delay profile, and the delay and Doppler indices are randomly generated within the range of $[0, l_{\mathrm{max}}]$ and $[-k_\mathrm{max}, k_\mathrm{max}]$. Specifically, we set $M=32$, $N=16$, $U=20$, $\Delta f=1.5\;\mathrm{kHz}$, $f_\mathrm{c}=4\;\mathrm{GHz}$, $l_{\mathrm{max}}=10$, and $k_\mathrm{max}=5$ \cite{li2021cross}, unless otherwise stated.
	
	We compare S1, S2, and S3 with the following benchmarks:
	\begin{itemize}
		\item The first benchmark is proposed in \cite{zhou2023otfs}, where the authors set the precoding matrix as $\boldsymbol{B}_u=(\boldsymbol{H}_u^H\boldsymbol{H}_u+\sigma^2\boldsymbol{I})^{-1}\boldsymbol{H}_u^H$, and $\hat{\boldsymbol{f}}$ is given by $\hat{\boldsymbol{f}}=\sum_{u=1}^{U}(\boldsymbol{H}_u\boldsymbol{B}_u\boldsymbol{x}_u+\boldsymbol{w}_u/\phi_u)$, where $\phi_u=\sqrt{P_\mathrm{t}/\mathrm{Tr}(\boldsymbol{B}_u\boldsymbol{B}_u^H)}$ is the power normalization factor. Then, MSE is calculated as
		\begin{align}\label{epsi}
			\epsilon=&\frac{1}{U^2}\left[\sum_{u=1}^{U}\mathrm{Tr}\left[\left(\boldsymbol{H}_u\boldsymbol{B}_u-\boldsymbol{I}\right)\left(\boldsymbol{H}_u\boldsymbol{B}_u-\boldsymbol{I}\right)^H\right]\right.\notag\\&\left.+\frac{\sigma^2\mathrm{Tr}(\boldsymbol{B}_u\boldsymbol{B}_u^H)}{P_t}\right].
	\end{align}
		\item The second benchmark \cite{zhou2023power} is proposed based on \cite{zhou2023otfs}, where the authors include power control at each device. In particular, MSE is calculated by replacing $P_\mathrm{t}$ with $p_u$, $p_u\leq P_\mathrm{t}$, in \eqref{epsi}, and optimal $p_u$ is obtained by exhavsive search to achieve the minimum MSE.
		\item The third benchmark is a simplified version of S3, where we only consider a precoding matrix at each device. By setting $\boldsymbol{V}^*=\boldsymbol{I}$ in \eqref{bu}, we obtain $\boldsymbol{B}_u^*$ as $\boldsymbol{B}_u^*=(\boldsymbol{H}_u^H\boldsymbol{H}_u+\lambda_u^*\boldsymbol{I})^{-1}\boldsymbol{H}_u^H$. MSE is obtained by setting $\boldsymbol{V}$ as $\boldsymbol{I}$ in \eqref{ems}.
		\item The fourth benchmark only considers a receiving filter matrix at FC. By setting $\boldsymbol{B}_u^*=\boldsymbol{I}$ in \eqref{v}, we obtain $\boldsymbol{V}^*$ as $\boldsymbol{V}^*=\left(\sum_{u=1}^{U}\boldsymbol{H}_u^H\right)\left(\sum_{u=1}^{U}\boldsymbol{H}_u\boldsymbol{H}_u^H+\sigma^2\boldsymbol{I}\right)^{-1}$. MSE is obtained by setting $\boldsymbol{B}_u$ as $\boldsymbol{I}$ in \eqref{ems}.
	\end{itemize}
	\subsection{Error Performance for S1, S2, and S3}
	
	\begin{figure}[!t]
		\begin{center}
			\includegraphics[width=0.9\columnwidth]{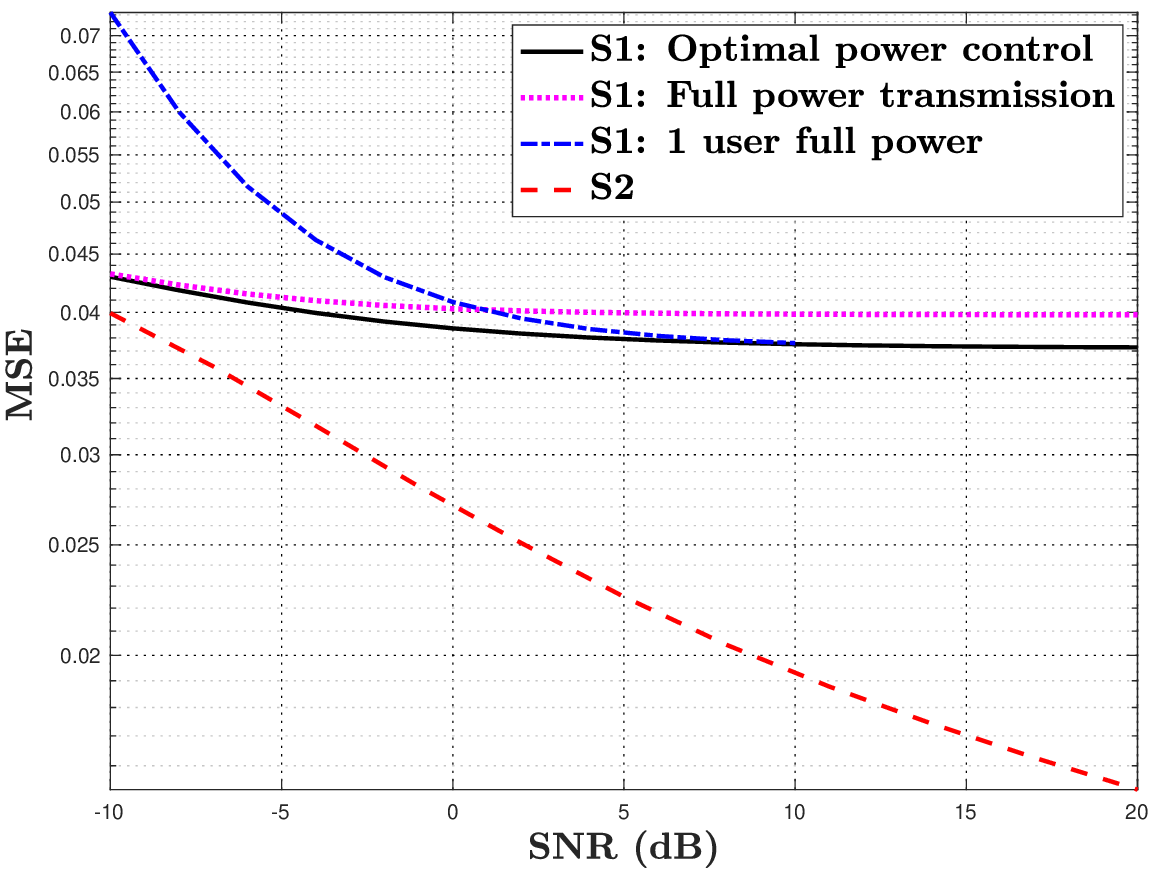}
			\caption{MSE versus SNR for S1 and S2, where different transmission power policies are applied.}\label{snr}\vspace{-0.5em}
		\end{center}\vspace{-4mm}
	\end{figure}
	
	In Fig. \ref{snr}, we plot MSE versus SNR for S1 and S2, where SNR is defined as $P_\mathrm{s}/\sigma^2$ and $P_\mathrm{s}$ is the average power budget. First, we apply different transmission power policies for S1. Our observations reveal that assigning optimal transmit power according to \eqref{pu} results in a significantly lower MSE compared to allotting full transmit power to all devices or to a singular user. This supports the optimal efficiency of $p_u^*$ and $\eta^*$ in minimizing MSE. Second, we observe that an MSE plateaus of S1 for larger SNR, which reveals the fact that interference plays the pivotal role in degrading the performance of estimation. We further observe that MSE for S2 is much lower than that of S1, which proves the superiority of ZP-assisted OTFS in mitigating the impact of interference. We also observe that this superiority becomes more obvious as SNR increases. We note that this superiority is established on the premise of sacrificing communication efficiency. Specifically, the reduction in transmission rate (measured in symbols per second) for ZP-assisted OTFS is given by $l_{u, \mathrm{max}}/T$.
	
	\begin{figure}[!t]
		\begin{center}
			\includegraphics[width=0.9\columnwidth]{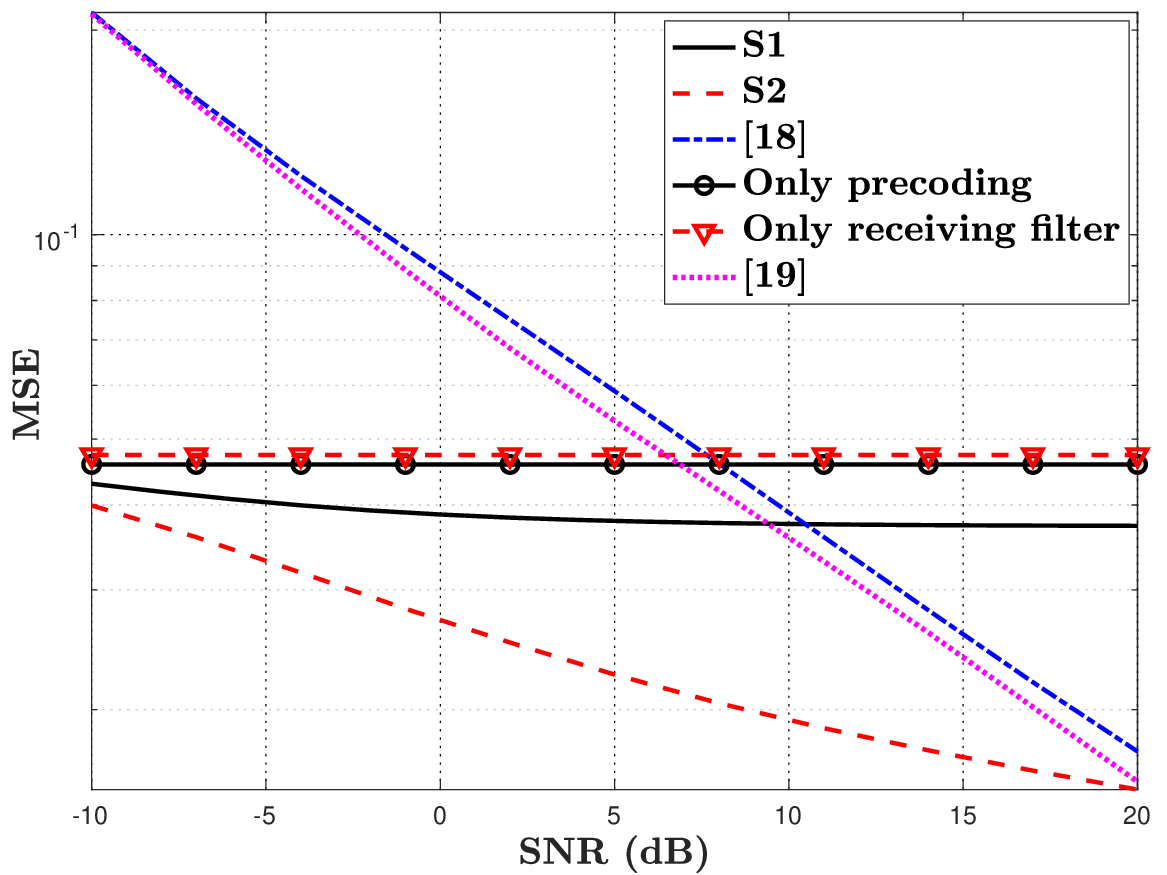}
			\caption{MSE versus SNR for S1, S2, and benchmarks.}\label{snr2}\vspace{-0.5em}
		\end{center}\vspace{-4mm}
	\end{figure}
	
	In Fig. \ref{snr2}, we compare the MSE for our proposed estimation methods with benchmarks. First, Fig. \ref{snr2} shows that our proposed estimation schemes S1 and S2 have a MSE superiority compared to the methods in \cite{zhou2023otfs,zhou2023power}, and this MSE superiority of S1 only holds for low SNR. Second, we observe that the MSE performance for our proposed methods is always better than that when only considering the precoding matrix at each device or only considering the receiving filter at the FC. This observation illustrates the necessity of considering both the pre-processing and post-processing at the TX and RX. Third, the MSEs for methods that consider only precoding or only the receiving filter remain nearly constant as the SNR changes. This occurs because using either precoding or a receiving filter alone fails to adequately compensate for signal distortion caused by channel fading, delay, and Doppler effects, resulting in errors significantly larger than those due to channel noise. This observation underscores the critical importance of incorporating both precoding and receiving filters to effectively recover the target function.
	\begin{figure}[!t]
		\begin{center}
			\includegraphics[width=0.9\columnwidth]{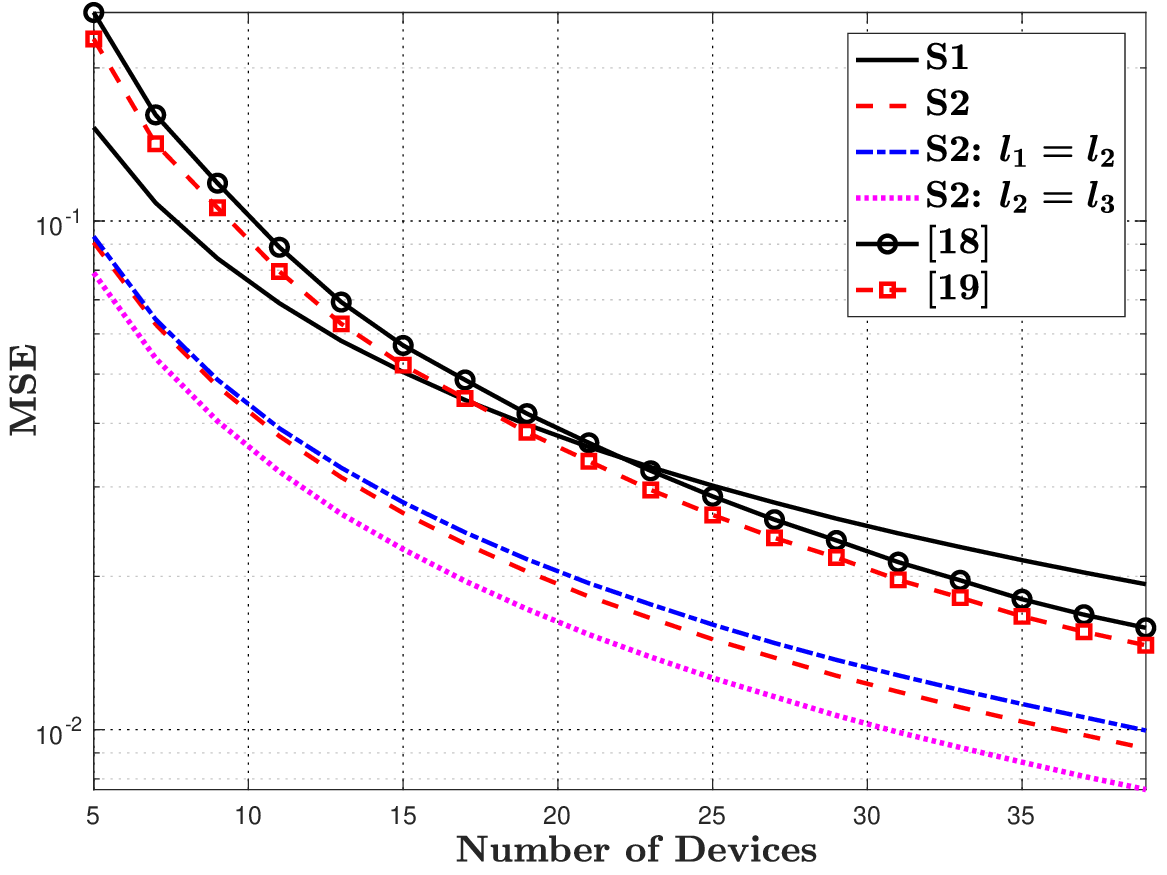}
			\caption{MSE versus number of devices, S1, S2, and the methods in \cite{zhou2023otfs,zhou2023power}. For S2, each path has a different delay or two paths have the same delay and $\mathrm{SNR}=10\;\mathrm{dB}$.}\label{devices}\vspace{-0.5em}
		\end{center}\vspace{-4mm}
	\end{figure}
	
	In Fig. \ref{devices}, we plot the MSE versus the number of devices participating in AirComp. First, we observe a reduction in MSE as the number of devices increases, aligning with the general understanding that increased device participation enhances estimation accuracy in AirComp. Second, we consider the scenario where two paths share the identical delay. We first assume $l_1=l_2\neq l_3\neq l_4$, which leads to two non-zero values in each row of $\boldsymbol{K}_{m,l_1}^u$. As $\boldsymbol{K}_{m,l_1}^u$ directly multiplies with the estimated function $\sum_{u=1}^{U}\boldsymbol{x}_{u,m}$, there exists interference that cannot be canceled during estimation, causing a slight rise in MSE compared to scenarios with varying delays in each path. We then assume $l_2=l_3\neq l_1\neq l_4$, which leads to two non-zero values in each row of $\boldsymbol{K}_{m,l_2}^u$. As $\boldsymbol{K}_{m,l_2}^u$ multiplies only with previously estimated interferences, effective interference cancellation occurs during the current estimation process. Moreover, the sharing of identical values by two paths enlarges the average pairwise differences between $l_1$, $l_2$, and $l_4$. Consequently, these two factors contribute to a lower MSE for $l_2=l_3$ compared to scenarios where each path experiences varying delays. Third, comparing the methods in \cite{zhou2023otfs,zhou2023power} with S1 and S2 reveals that the methods in \cite{zhou2023otfs,zhou2023power} outperform S1 when the number of devices is large, while S2 consistently achieves a lower MSE than the benchmarks in \cite{zhou2023otfs,zhou2023power}.
	\begin{figure}[!t]
		\begin{center}
			\includegraphics[width=0.9\columnwidth]{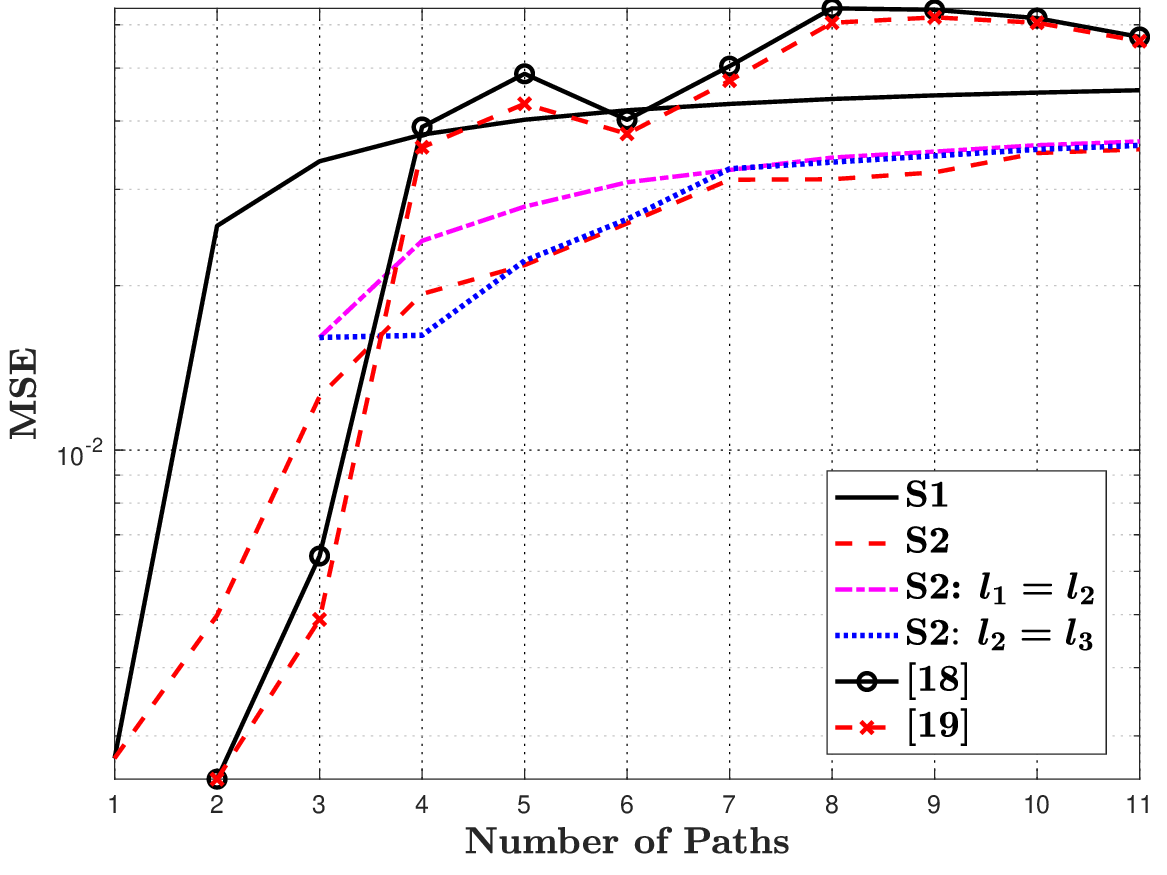}
			\caption{MSE versus number of paths for S1, S2, and methods in \cite{zhou2023otfs,zhou2023power}, where $\mathrm{SNR}=10\;\mathrm{dB}$.}\label{path}\vspace{-0.5em}
		\end{center}\vspace{-4mm}
	\end{figure}
	
	In Fig. \ref{path}, we plot MSE versus the number of paths for S1, S2, and benchmarks in \cite{zhou2023otfs,zhou2023power}. For S2, we consider the scenarios that either each path has a different delay or two paths share the same delay. For different delay in each path, the number of paths is from 1 to 11. for two paths sharing the same delay, the number of paths is from 3 to 11. First, we observe that the MSEs for S1 and S2 increase as the number of paths increases. This is due to the fact that a larger number of paths introduces more interference terms. Second, the MSE gap between S1 and S2 decreases as the number of paths increases, which indicates that the SIC-based algorithm in S2 is more proficient in mitigating the interference when the number of channel paths is small. Third, for the scenario that two paths share the same delay, we consider either $l_1=l_2$ or $l_2=l_3$. We observe that the MSE for $l_1=l_2$ is always larger than that for each path having a different delay. Eventually, when the number of paths is sufficiently large, the MSE performance is the same for scenarios where $l_1=l_2$, $l_2=l_3$, and each path has a different delay. Fourth, the MSEs of S1 and S2 are lower than those of \cite{zhou2023otfs,zhou2023power} when the number of paths exceeds 3, demonstrating the superiority of our proposed methods in handling a large number of channel paths.
	
	\begin{figure}[!t]
		\begin{center}
			\includegraphics[width=0.9\columnwidth]{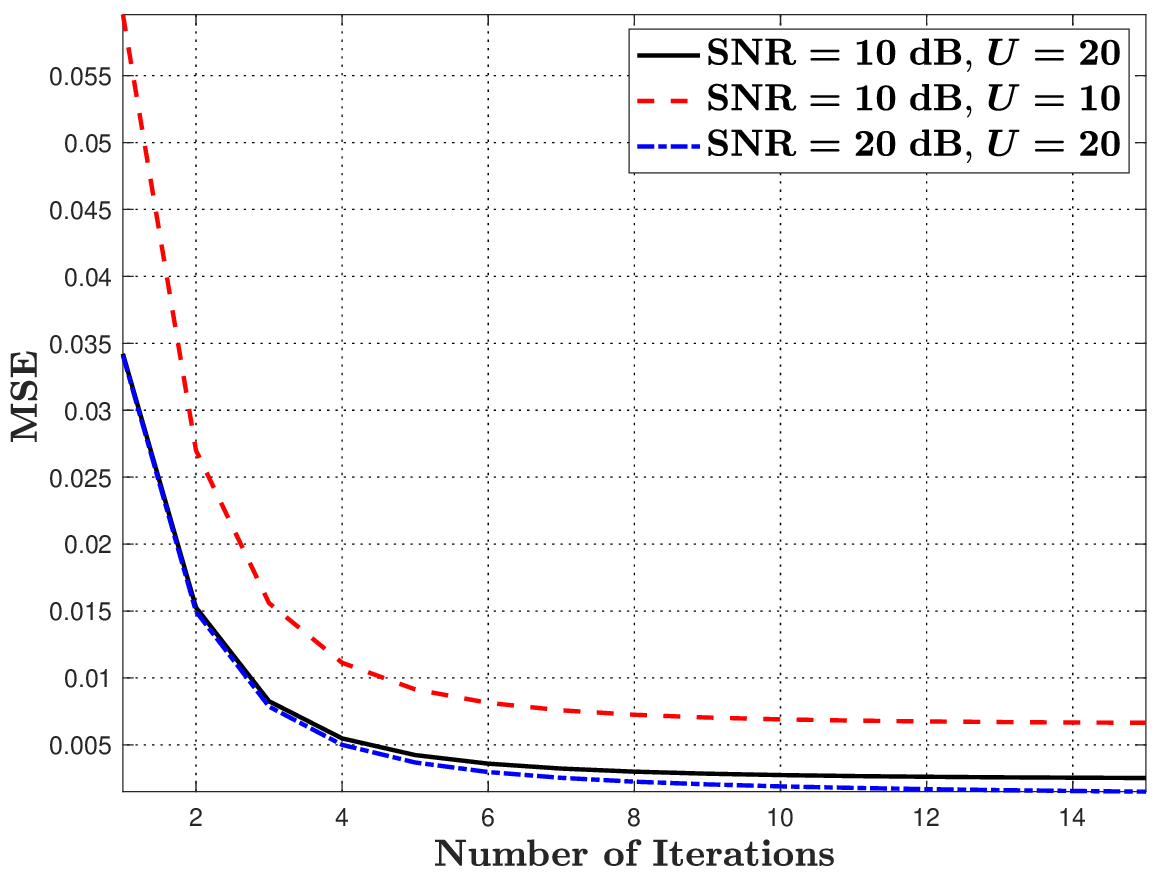}
			\caption{MSE versus the number of iterations for S3 with varying SNR and $U$, where the number of iterations is 10 and $\mathrm{SNR}=10\;\mathrm{dB}$.}\label{noi}\vspace{-0.5em}
		\end{center}\vspace{-4mm}
	\end{figure}
	In Fig. \ref{noi}, we evaluate the MSE performance for S3. First, we observe that the MSE decreases with the increase in the number of iterations in the algorithm, and the MSE eventually converges when the number of iterations is sufficiently large (around 10). Second, we observe that the MSE performance is better for a higher value of SNR and a larger number of participating devices. The performance superiority of high SNR becomes more obvious with the increase in the number of iterations. 
	
	\begin{figure}[!t]
		\begin{center}
			\includegraphics[width=0.9\columnwidth]{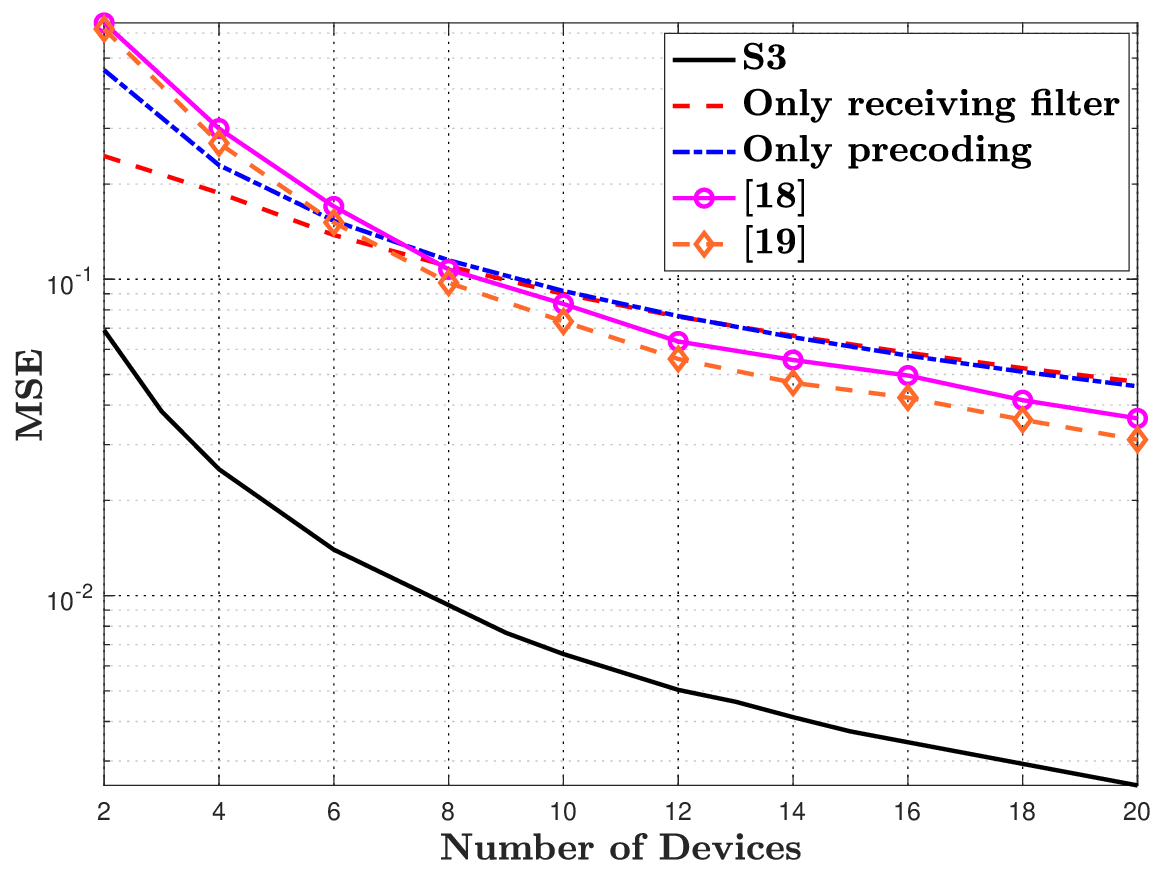}
			\caption{MSE versus the number of devices for S3 and benchmarks.}\label{de}\vspace{-0.5em}
		\end{center}\vspace{-4mm}
	\end{figure}
	In Fig. \ref{de}, we compare the error performance of the proposed iterative algorithm in S3 with bechmarks, where the number of iterations is chosen as 10. From Fig. \ref{de}, we clearly observe that S3 achieves a much better error performance over different number of devices than the methods in \cite{zhou2023otfs,zhou2023power}. Moreover, comparing to Fig. \ref{devices}, the error performance of S3 is also better than that of S1 and S2.
	
	\begin{figure}[!t]
		\begin{center}
			\includegraphics[width=0.9\columnwidth]{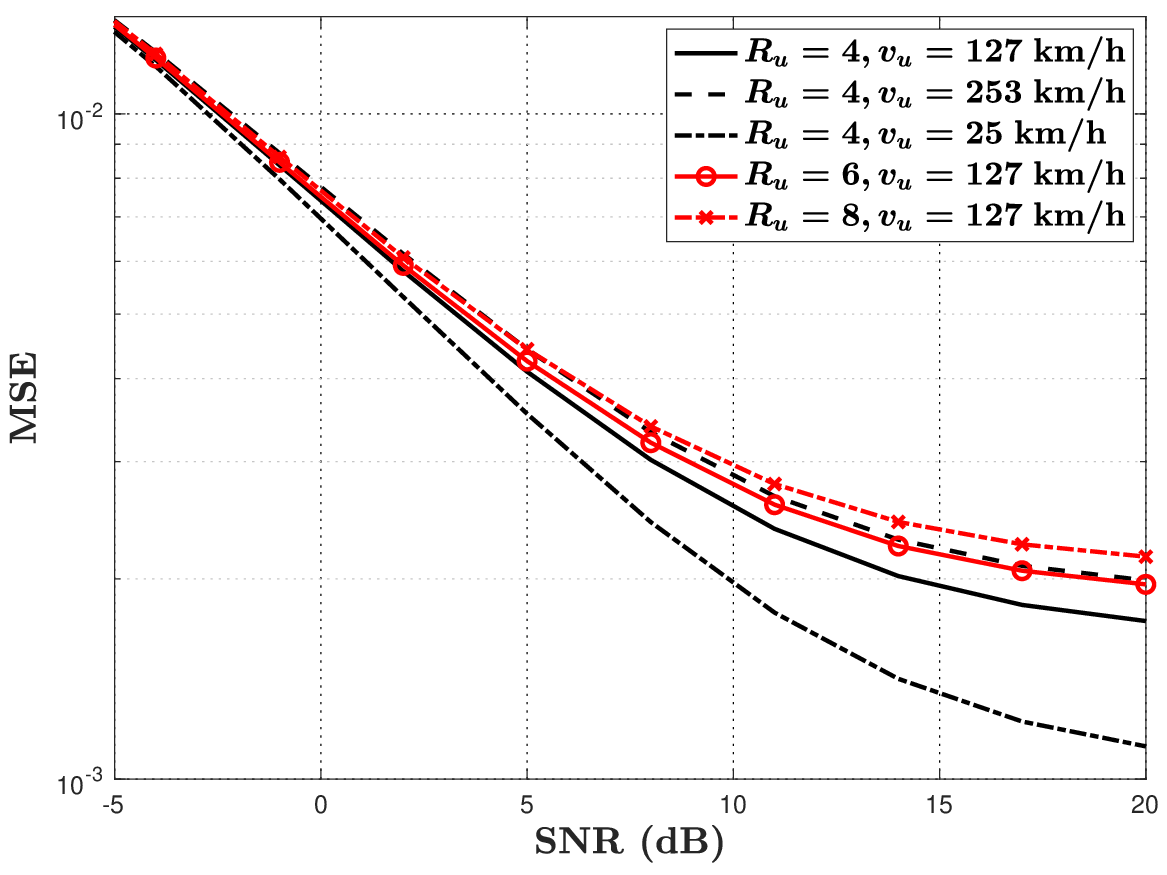}
			\caption{MSE versus SNR for S3 with varying $R_u$ and $v_u$.}\label{sn}\vspace{-0.5em}
		\end{center}\vspace{-4mm}
	\end{figure}
	In Fig. \ref{sn}, we evaluate the error performance of S3 for different number of channel paths and the moving speed of the devices. For each $v_u$, we calculate $k_\mathrm{max}$ as $k_\mathrm{max}=v_uf_\mathrm{c}N/(c\Delta f)$. Then, the Doppler index of each path is randomly generated within the range $[-k_\mathrm{max}, k_\mathrm{max}]$. First, we observe that the MSE increases with the increase in the moving speed of devices. This is due to the fact that a higher moving speed leads to a higher Doppler shift which degrades the estimation performance at FC. Second, when $\mathrm{SNR}=20\;\mathrm{dB}$, we observe a more significant increase in MSE when the velocity $v_u$ rises from $25\;\mathrm{km}/\mathrm{h}$ to $127\;\mathrm{km}/\mathrm{h}$ compared to the increase from $127\;\mathrm{km}/\mathrm{h}$ to $253\;\mathrm{km}/\mathrm{h}$. This observation suggests that OTFS demonstrates greater robustness to velocity changes in the high-speed range. Third, we observe that the MSE improves with the increase in the number of channel paths between each device and FC. This is because a larger number of paths introduces more ISI and ILI. 
	\subsection{Computational Complexity}
	\begin{figure*}[!t]
		\centering
		
		\subfigure[Running time versus $M\times N$]{
			\begin{minipage}[t]{0.32\linewidth}
				\centering
				\includegraphics[width=0.95\columnwidth]{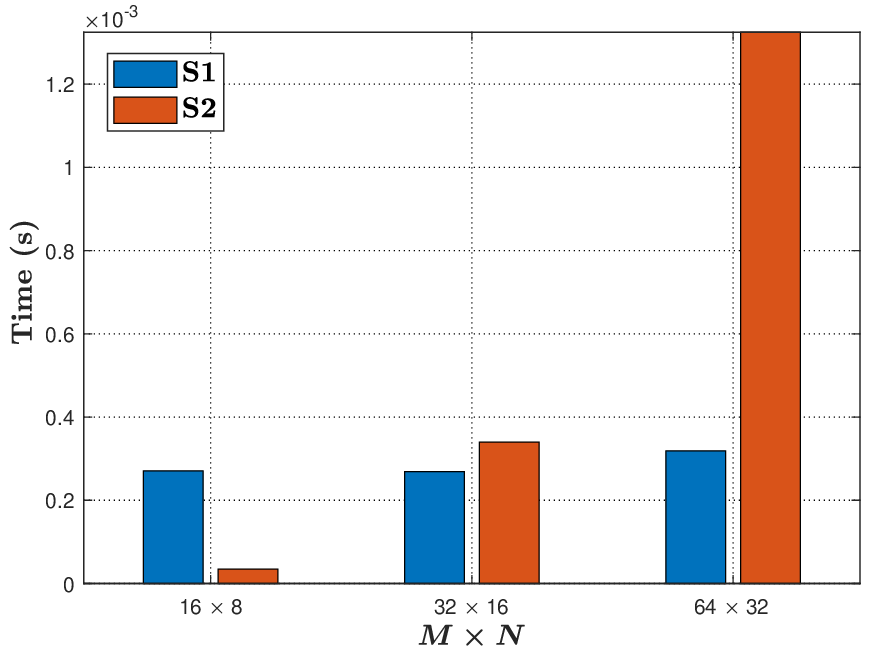}
				\label{a1}
			\end{minipage}%
		}%
		\subfigure[Running time versus $M\times N$]{
			\begin{minipage}[t]{0.32\linewidth}
				\centering
				\includegraphics[width=0.95\columnwidth]{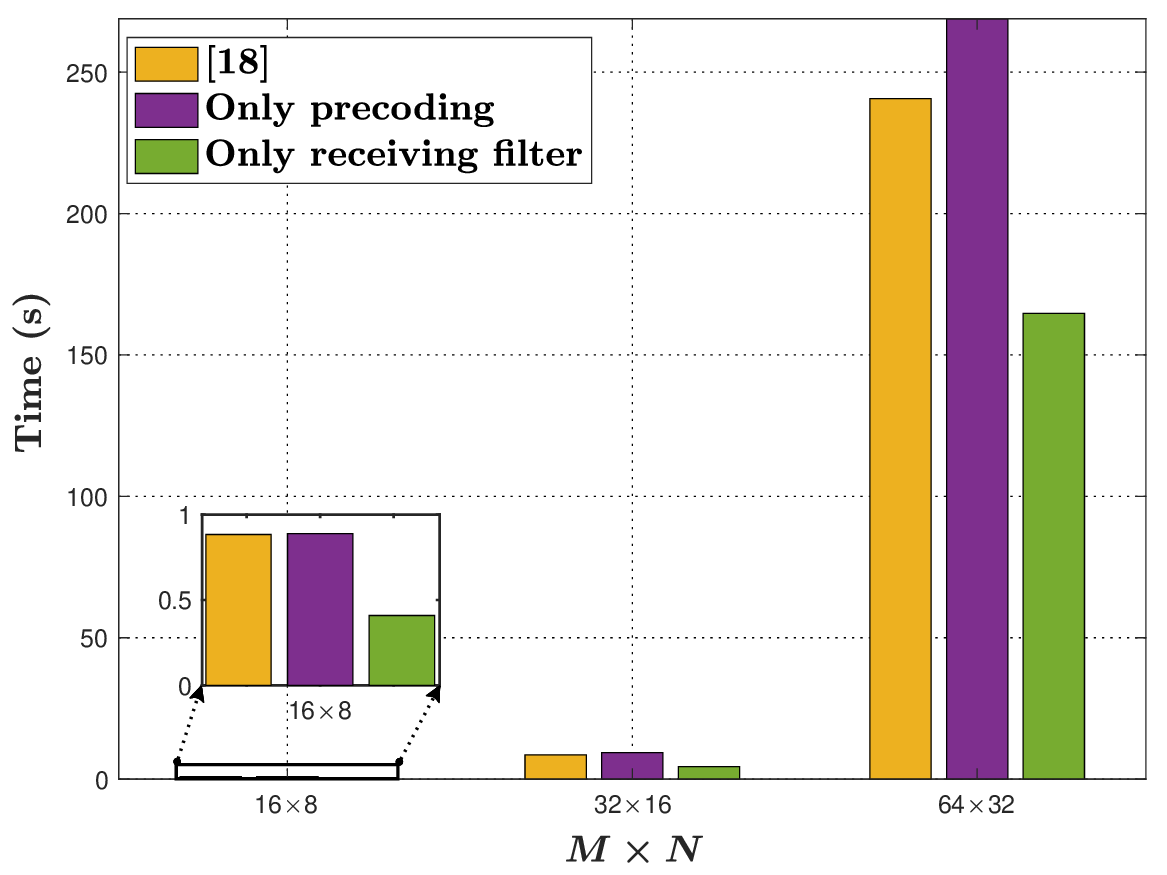}
				\label{a2}
			\end{minipage}%
		}
		\subfigure[Running time versus $M\times N$]{
			\begin{minipage}[t]{0.32\linewidth}
				\centering
				\includegraphics[width=0.95\columnwidth]{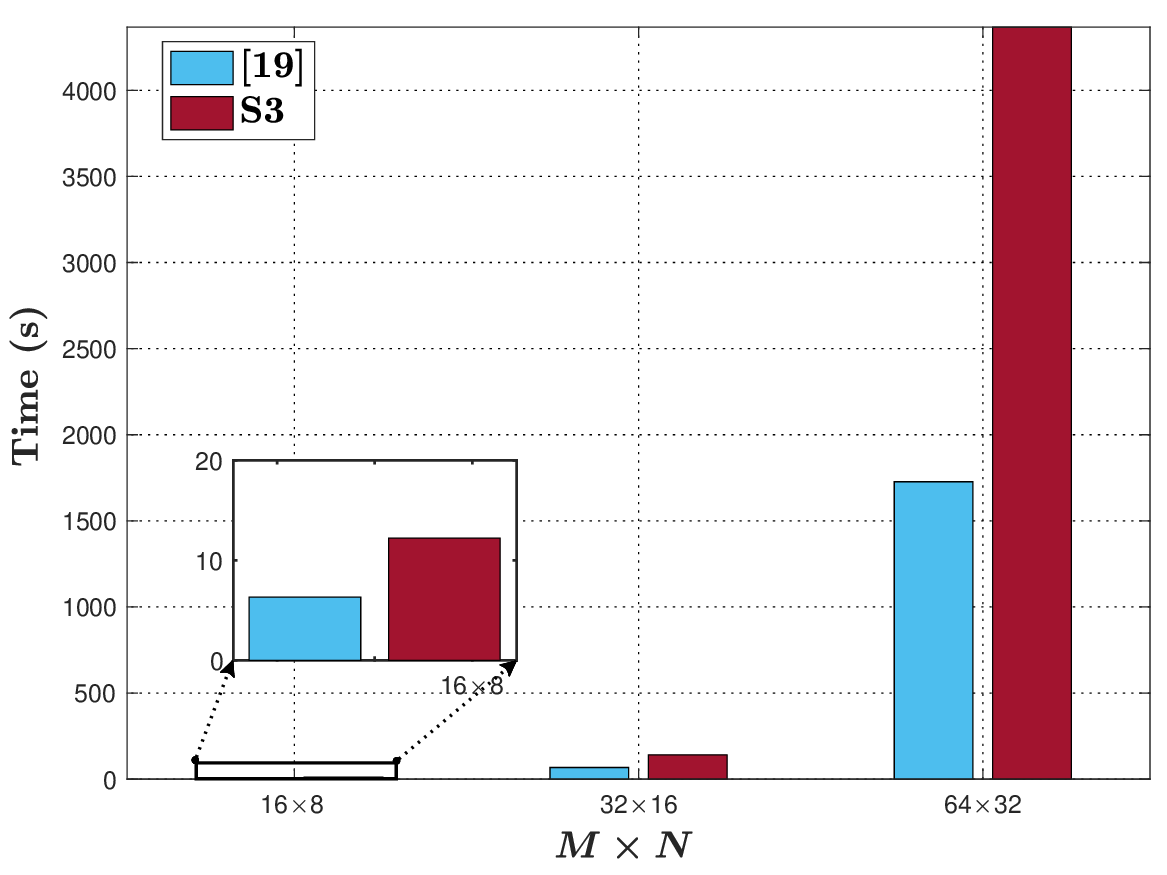}
				\label{a3}
			\end{minipage}%
		}
		\quad
		\subfigure[Running time versus $R_u$]{
			\begin{minipage}[t]{0.32\linewidth}
				\centering
				\includegraphics[width=0.95\columnwidth]{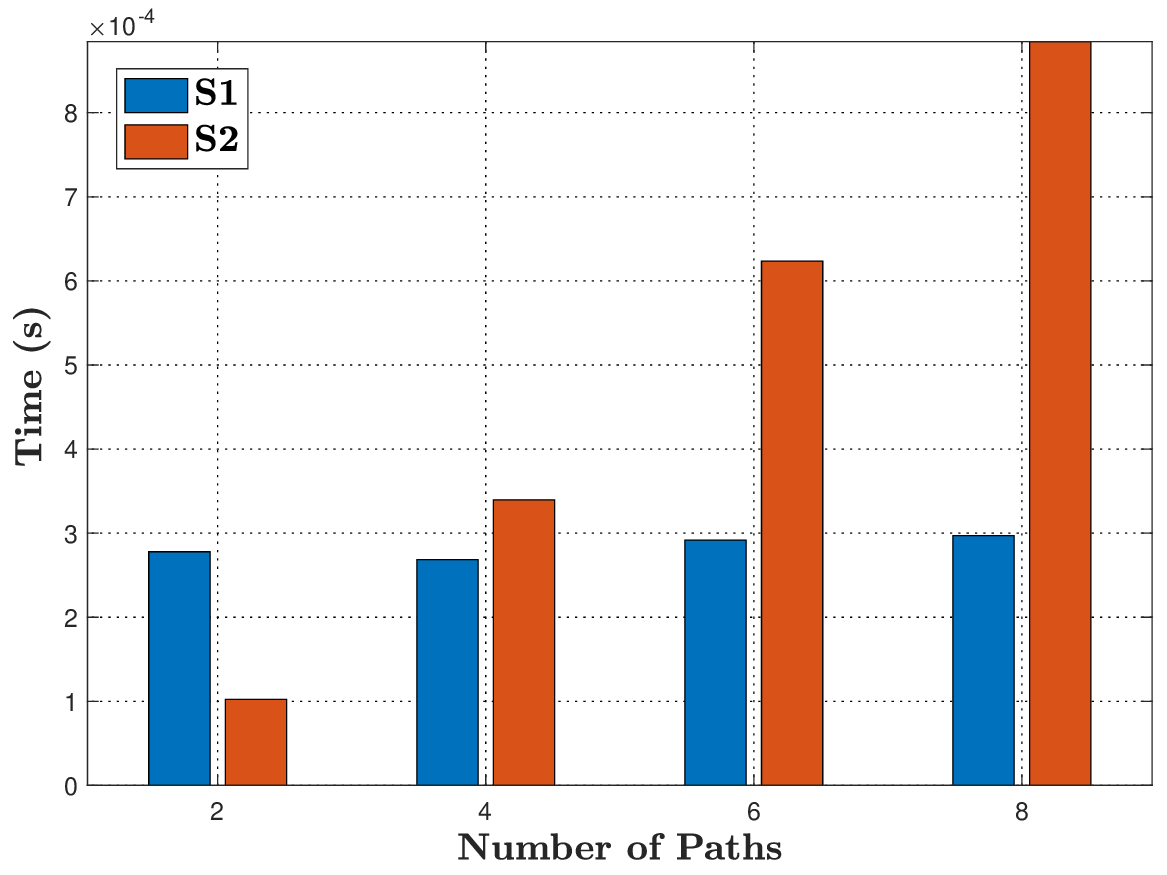}
				\label{a4}
			\end{minipage}%
		}%
		\subfigure[Running time versus $R_u$]{
			\begin{minipage}[t]{0.32\linewidth}
				\centering
				\includegraphics[width=0.95\columnwidth]{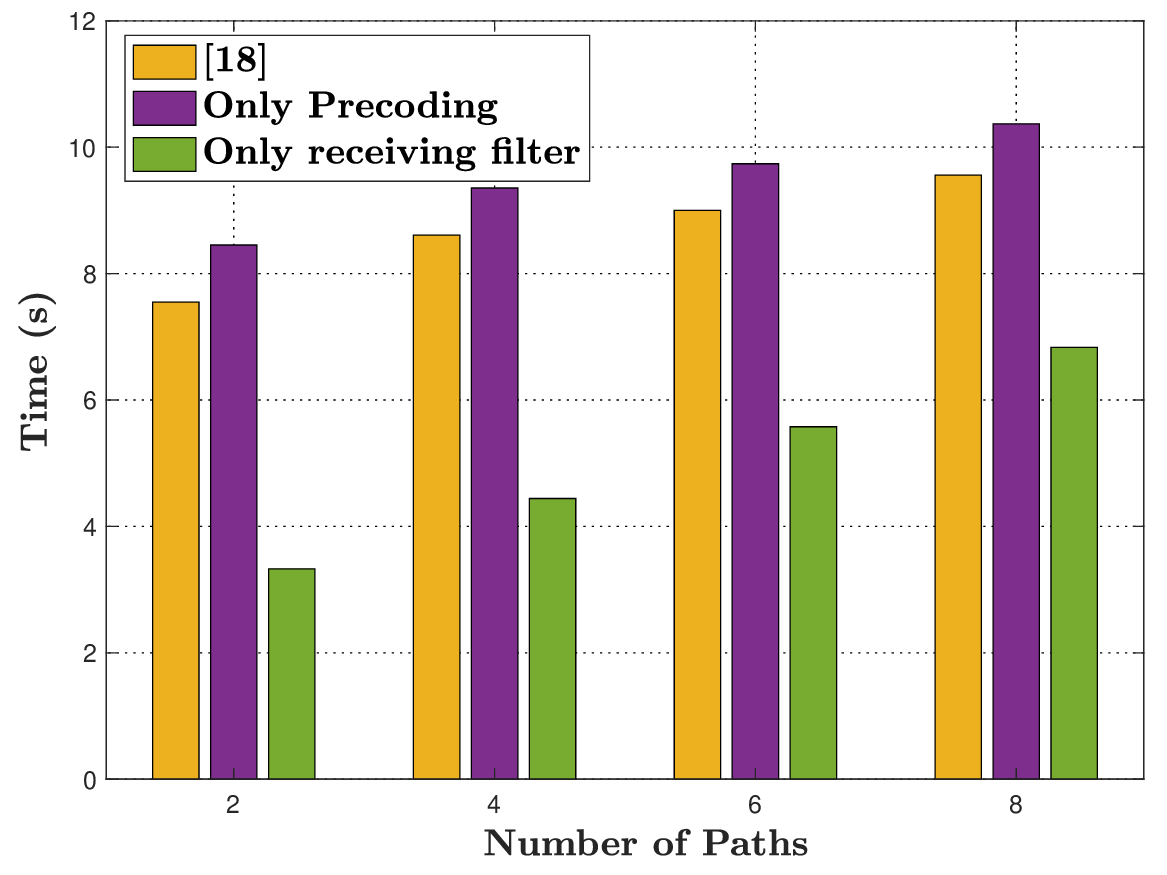}
				\label{a5}
			\end{minipage}%
		}
		\subfigure[Running time versus $R_u$]{
			\begin{minipage}[t]{0.32\linewidth}
				\centering
				\includegraphics[width=0.95\columnwidth]{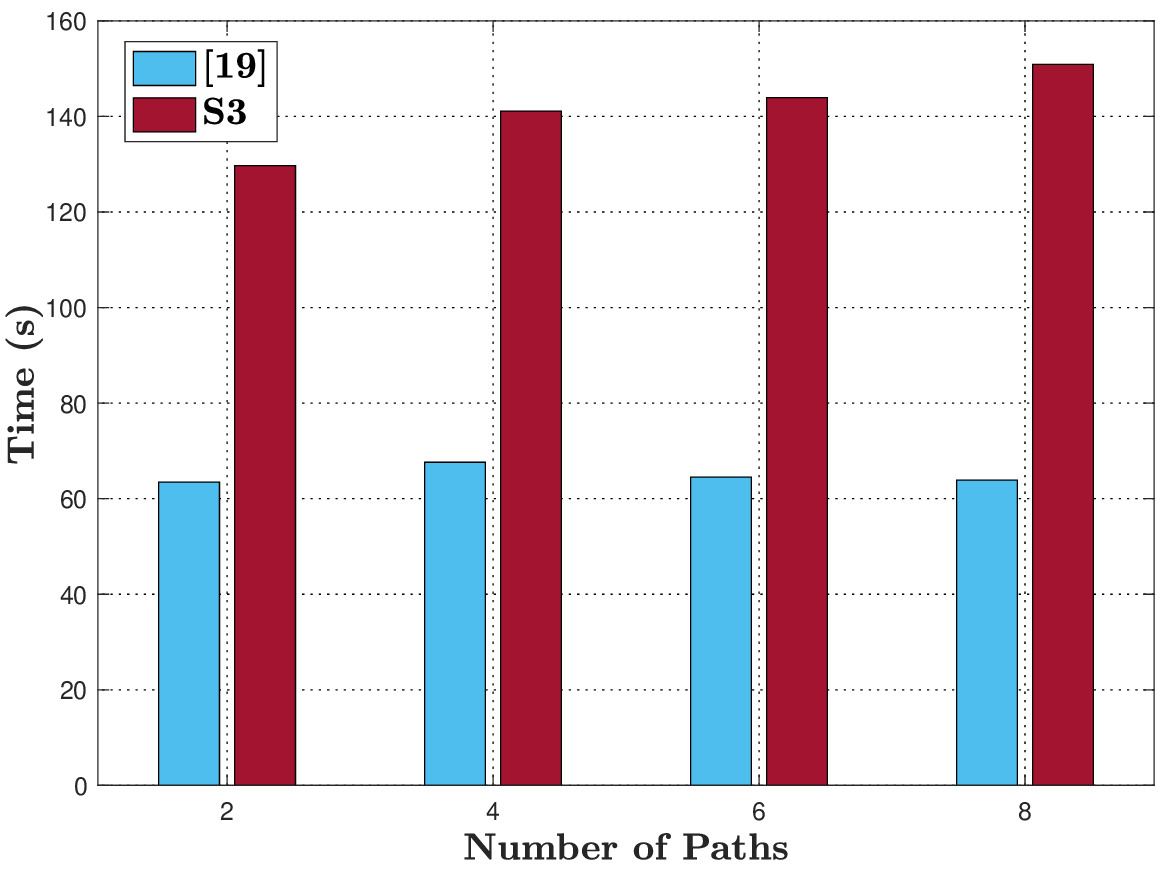}
				\label{a6}
			\end{minipage}%
		}
		\centering
		\caption{Running time versus $M\times N$ for (a): S1 and S2, (b): the method in \cite{zhou2023otfs} and the method that only applying precoding or receiving filter, and (c): S3 and the method in \cite{zhou2023power}. Running time versus the number of channel paths for (d): S1 and S2, (e): the method in \cite{zhou2023otfs} and the method that only applying precoding or receiving filter, and (f): S3 and the method in \cite{zhou2023power}.}
		\label{ame}
	\end{figure*}
	In Fig. \ref{ame}, we evaluate the computational complexity for S1, S2, S3, and the benchmarks. In particular, we record the running time of each realization using Matlab for S1, S2, S3, and benchmarks and use this time as a criterion to reflect the computational complexity. We note that all the experiments are conducted within a consistent computer environment. In Figs. \ref{a1}, \ref{a2}, and \ref{a3}, we plot the running time versus $M\times N$. Here, $M\times N$ represents the size of one OTFS frame. In Figs. \ref{a4}, \ref{a5}, and \ref{a6}, we plot the running time versus the number of channel paths. First, we observe that the running time of S1 and S2 is much less than that of the benchmarks and S3. In particular, when $M=64$ and $N=32$, the running time of the methods in \cite{zhou2023otfs,zhou2023power} is $10^5$ and $10^6$ times longer, respectively, than the proposed S1 and S2 in this paper. Therefore, compared to benchmarks, S2 can not only achieve a better error performance, but also have a much lower computational complexity. Second, we observe that the complexity of S2, S3, and benchmarks increases with the increase in $M\times N$, and the increase in benchmarks and S3 is much faster than that in S2. Third, as illustrated in Fig. \ref{ame}, the complexity of S2, the method in \cite{zhou2023otfs}, and S3 increases with the increase in the number of channel paths, and the rate of this increase in S2 is much faster than that in \cite{zhou2023otfs} and S3. Fourth, our observations reveal that S1 consistently maintains a low level of complexity, which exhibits remarkable robustness against variations in both $M×N$ and the number of channel paths.
	
	In summary, each scheme offers unique advantages. S1 excels in simplicity, low computational complexity, and robustness to varying OTFS size and the number of channel paths. S2 strikes a balance with better error performance than S1 and lower complexity than S3. S3 offers the best error performance but at the cost of higher computational complexity.Ultimately, there is a trade-off between computational complexity and error performance, and the choice of scheme depends on the specific requirements of the scenario.
	
	\section{Conclusion}\label{c}
	In this paper, we considered an OTFS-based AirComp over double-selective time-varying channels, where mobile devices first modulate their data on the delay-Doppler (DD) domain and then transmit them to the fusion center (FC) for averaging. In particular, we proposed three schemes S1, S2, and S3 of increasing complexity for the target function estimation. In S1, we derived the closed-form expressions for the optimal transmit power and denoising factor. In S2, we applied a ZP-assisted OTFS-based AirComp, where we proposed an SIC-based algorithm and derived the closed-form expressions for the optimal transmit power, denoising factor, and the weight  for interference cancellation. In S3, we proposed an iterative algorithm to jointly optimize the precoding matrix and receiving filter matrix. Numerical results validated the superiority of the proposed schemes in the error performance than benchmarks. They also showed a low computational complexity of S1 and S2 compared to benchmarks. Future research directions include conducting real-world OTFS measurements in vehicular communication scenarios to evaluate the practical implementation of the proposed algorithms. Additionally, given the shared parameters between OTFS and radar sensing, such as delay and Doppler, the application of OTFS in Integrated Sensing and Communication (ISAC) has garnered considerable interest. Building on the results of this paper, an intriguing area of investigation emerges in exploring an integrated approach to sensing, communication, and computation within the DD domain.
	
	\appendices
	\section{Proof of Theorem \ref{t1}}\label{A1}
	We note that \eqref{pu} can be directly derived from \eqref{p} and \eqref{I}. For \eqref{e}, we need to prove $\eta_u^*=\hat{\eta}_u$ when $\eta\in I_{u^*}$, which is equvalient to confirm that $\eta_u^*\neq P_\mathrm{s}(\sum_{i=1}^{R_{u^*+1}}|h_{u^*+1,i}|^2/|h_{u^*+1, 1}|)^2$ and $\eta_u^*\neq P_\mathrm{s}(\sum_{i=1}^{R_u^*}|h_{u^*,i}|^2/|h_{u^*, 1}|)^2$ based on \eqref{et}.
	
	We first demonstrate $\eta_u^*\neq P_\mathrm{s}(\sum_{i=1}^{R_u^*}|h_{u^*,i}|^2/|h_{u^*, 1}|)^2$ by using the proof of contradiction. If $\eta_u^*=P_\mathrm{s}(\sum_{i=1}^{R_u^*}|h_{u^*,i}|^2/|h_{u^*, 1}|)^2$, $P_\mathrm{s}(\sum_{i=1}^{R_u^*}|h_{u^*,i}|^2/|h_{u^*, 1}|)^2$ becomes the global optimal and thus is the optimal for $H_{u^*-1}(\eta)$ and $H_{u^*}(\eta)$. Therefore, according to \eqref{et}, we have 
	\begin{align}\label{bpl}
		P_\mathrm{s}\!\!\left(\frac{\sum_{i=1}^{R_u^*}|h_{u^*,i}|^2}{|h_{u^*, 1}|}\right)^2\leq\hat{\eta}_{u^*-1}, P_\mathrm{s}\left(\frac{\sum_{i=1}^{R_u^*}|h_{u^*,i}|^2}{|h_{u^*, 1}|}\right)^2\geq\hat{\eta}_{u^*}.
	\end{align}
	Next, we derive $P_\mathrm{s}(\sum_{i=1}^{R_u}|h_{u,i}|^2/|h_{u, 1}|)^2-\hat{\eta}_u$ as 
	\begin{align}\label{b}
		&P_\mathrm{s}\left(\frac{\sum_{i=1}^{R_u}|h_{u,i}|^2}{|h_{u, 1}|}\right)^2-\hat{\eta}_u=\left(\frac{\sqrt{P_\mathrm{s}}\sum_{i=1}^{R_u}|h_{u,i}|^2}{|h_{u,1}|}+\sqrt{\hat{\eta}_u}\right)\notag\\&\times\left(\frac{F(u)}{\sqrt{P_\mathrm{s}}|h_{u,1}|\sum_{j=1}^{u}|h_{j,1}|}\right),
	\end{align}
	where $F(u)=P_\mathrm{s}\sum_{j=1}^{u-1}|h_{j,1}|\sum_{i=1}^{R_u}|h_{u,i}|^2-P_\mathrm{s}|h_{u,1}|\sum_{j=1}^{u-1}\sum_{i=1}^{R_j}|h_{j,i}|^2-|h_{u,1}|\sigma^2$. We then calculate $P_\mathrm{s}(\sum_{i=1}^{R_u}|h_{u,i}|^2/|h_{u, 1}|)^2-\hat{\eta}_{u-1}$ as
	\begin{align}\label{bp}
		&P_\mathrm{s}\!\left(\!\frac{\sum_{i=1}^{R_u}|h_{u,i}|^2}{|h_{u, 1}|}\!\right)^2\!-\!\hat{\eta}_{u-1}\!=\! \left(\frac{\sqrt{P_\mathrm{s}}\sum_{i=1}^{R_u}|h_{u,i}|^2}{|h_{u,1}|}+\sqrt{\hat{\eta}_{u-1}}\right)\notag\\&\times\left(\frac{F(u)}{\sqrt{P_\mathrm{s}}|h_{u,1}|\sum_{j=1}^{u}|h_{j,1}|}\right).
	\end{align}
	From \eqref{b} and \eqref{bp}, we can see that $P_\mathrm{s}(\sum_{i=1}^{R_u}|h_{u,i}|^2/|h_{u, 1}|)^2-\hat{\eta}_u$ and $P_\mathrm{s}(\sum_{i=1}^{R_u}|h_{u,i}|^2/|h_{u, 1}|)^2-\hat{\eta}_{u-1}$ have the same sign, which is contradict to \eqref{bpl}. Therefore, we have $\eta_u^*\neq P_\mathrm{s}(\sum_{i=1}^{R_u^*}|h_{u^*,i}|^2/|h_{u^*, 1}|)^2$. We can also prove $\eta_u^*\neq P_\mathrm{s}(\sum_{i=1}^{R_{u^*+1}}|h_{u^*+1,i}|^2/|h_{u^*+1, 1}|)^2$ by using a similar method, which is omitted here due to the page limit.
	
	\section{Proof of Theorem \ref{t2}}\label{A2}
	We first derive $\zeta_{1,0}^*$. From \eqref{eps}, we can see that only $\E[G(\zeta_{1,0})^2]$ relates to $\zeta_{1,0}$. Therefore, we need to find $\zeta_{1,0}^*$ that minimizes $\E[G(\zeta_{1,0})^2]$. We note that the expectation of $G(\zeta_{1,0})$ is taken over $x_{u,\gamma_0}$, $w_{\gamma_1}$, and $w_{\gamma_0}$. As $x_{u,\gamma_0}$, $w_{\gamma_1}$, and $w_{\gamma_0}$ are independent with each other, $\E[G(\zeta_{1,0})^2]$ is derived as 
	\begin{align}\label{egd}
		&\E[G(\zeta_{1,0})^2]=\frac{\sum_{u=1}^{U}p_{u,0}^*|h_{u,1}|^2+\sigma^2}{\eta_0^*}\zeta_{1,0}^2\notag\\&\!-\!\frac{2\sum_{u=1}^{U}|h_{u,2}||h_{u,1}|p_{u,0}^*}{\sqrt{\eta_0^*}}\zeta_{1,0}\!+\!\!\sum_{u=1}^{U}|h_{u,2}|^2p_{u,0}^*\!+\!\sigma^2.
	\end{align}
	Since \eqref{egd} is a quadratic function, the optimal $\zeta_{1,0}^*$ can be obtained by calculating $\partial\E[G(\zeta_{1,0})^2]/\partial\zeta_{1,0}=0$, which leads to \eqref{d}. By substituting $\zeta_{1,0}^*$ into \eqref{egd}, we can obtain $\min\E[G(\zeta_{1,0})^2]$. After obtaining $\zeta_{1,0}^*$, the optimization problem in \eqref{eps} becomes 
	\begin{align}\label{mpug}
		\min_{p_{u,1}\geq0, \eta_1\geq0}  &\epsilon_1=\sum_{u=1}^{U}\left(\frac{p_{u,1}|h_{u,1}|}{\sqrt{\eta_1}}-1\right)^2+\frac{\min\E[G(\zeta_{1,0})^2]}{\eta_1}, \notag\\
		&\mathrm{s.t.}\;p_{u,1}\leq P_\mathrm{s}.
	\end{align}
	We note that the optimization problem in \eqref{mpug} is the same as that in \cite{cao2020optimized}. Then, \eqref{pug1} and \eqref{egl} can be obtained following \cite[Eqs. (24), (25)]{cao2020optimized}, whose derivation processes are omitted here due to the page limit.
	\bibliographystyle{IEEEtran}
	\bibliography{ref}

\begin{thebibliography}{10}
\providecommand{\url}[1]{#1}
\csname url@samestyle\endcsname
\providecommand{\newblock}{\relax}
\providecommand{\bibinfo}[2]{#2}
\providecommand{\BIBentrySTDinterwordspacing}{\spaceskip=0pt\relax}
\providecommand{\BIBentryALTinterwordstretchfactor}{4}
\providecommand{\BIBentryALTinterwordspacing}{\spaceskip=\fontdimen2\font plus
\BIBentryALTinterwordstretchfactor\fontdimen3\font minus
  \fontdimen4\font\relax}
\providecommand{\BIBforeignlanguage}[2]{{%
\expandafter\ifx\csname l@#1\endcsname\relax
\typeout{** WARNING: IEEEtran.bst: No hyphenation pattern has been}%
\typeout{** loaded for the language `#1'. Using the pattern for}%
\typeout{** the default language instead.}%
\else
\language=\csname l@#1\endcsname
\fi
#2}}
\providecommand{\BIBdecl}{\relax}
\BIBdecl

\bibitem{huang2022}
X.~Huang, H.~Hellström, and C.~Fischione, ``Interference cancellation for
  {OTFS}-based over-the-air computation,'' in \emph{Proc. IEEE ICC Workshop
  2024}, Denver, USA, Jun. 2024, pp. 1--6.

\bibitem{nguyen20216g}
D.~C. Nguyen, M.~Ding, P.~N. Pathirana, A.~Seneviratne, J.~Li, D.~Niyato,
  O.~Dobre, and H.~V. Poor, ``{6G} internet of things: A comprehensive
  survey,'' \emph{IEEE Internet Things J.}, vol.~9, no.~1, pp. 359--383, Jan.
  2021.

\bibitem{delltec}
\BIBentryALTinterwordspacing
Dell, ``Edge to core and the internet of things.'' [Online]. Available:
  \url{https://infohub.delltechnologies.com/t/edge-to-core-and-the-internet-of-things-2/}
\BIBentrySTDinterwordspacing

\bibitem{csahin2023survey}
A.~{\c{S}}ahin and R.~Yang, ``A survey on over-the-air computation,''
  \emph{IEEE Commun. Surv. Tutor.}, vol.~25, no.~3, pp. 1877--1908, 3rd Quart.
  2023.

\bibitem{wang2022edge}
S.~Wang, Y.~Hong, R.~Wang, Q.~Hao, Y.-C. Wu, and D.~W.~K. Ng, ``Edge federated
  learning via unit-modulus over-the-air computation,'' \emph{IEEE Trans.
  Commun.}, vol.~70, no.~5, pp. 3141--3156, May 2022.

\bibitem{nazer2007computation}
B.~Nazer and M.~Gastpar, ``Computation over multiple-access channels,''
  \emph{IEEE Trans. Inf. Theory}, vol.~53, no.~10, pp. 3498--3516, Oct. 2007.

\bibitem{gastpar2008uncoded}
M.~Gastpar, ``Uncoded transmission is exactly optimal for a simple {Gaussian}
  “sensor” network,'' \emph{IEEE Trans. Inf. Theory}, vol.~54, no.~11, pp.
  5247--5251, Nov. 2008.

\bibitem{yang2020federated}
K.~Yang, T.~Jiang, Y.~Shi, and Z.~Ding, ``Federated learning via over-the-air
  computation,'' \emph{IEEE Trans. Wireless Commun.}, vol.~19, no.~3, pp.
  2022--2035, Mar. 2020.

\bibitem{shao2021federated}
Y.~Shao, D.~G{\"u}nd{\"u}z, and S.~C. Liew, ``Federated edge learning with
  misaligned over-the-air computation,'' \emph{IEEE Trans. Wirel. Commun.},
  vol.~21, no.~6, pp. 3951--3964, Jun. 2021.

\bibitem{cao2020optimized}
X.~Cao, G.~Zhu, J.~Xu, and K.~Huang, ``Optimized power control for over-the-air
  computation in fading channels,'' \emph{IEEE Trans. Wirel. Commun.}, vol.~19,
  no.~11, pp. 7498--7513, Nov. 2020.

\bibitem{zhu2019broadband}
G.~Zhu, Y.~Wang, and K.~Huang, ``Broadband analog aggregation for low-latency
  federated edge learning,'' \emph{IEEE Trans. Wirel. Commun.}, vol.~19, no.~1,
  pp. 491--506, Jan. 2020.

\bibitem{tegin2023federated}
B.~Tegin and T.~M. Duman, ``Federated learning with over-the-air aggregation
  over time-varying channels,'' \emph{IEEE Trans. Wirel. Commun.}, vol.~22,
  no.~8, Aug. 2023.

\bibitem{hadani2017orthogonal}
R.~Hadani \emph{et~al.}, ``Orthogonal time frequency space modulation,'' in
  \emph{IEEE WCNC 2017}, San Francisco, CA, USA, Mar. 2017, pp. 1--6.

\bibitem{raviteja2018interference}
P.~Raviteja \emph{et~al.}, ``Interference cancellation and iterative detection
  for orthogonal time frequency space modulation,'' \emph{IEEE Trans. Wirel.
  Commun.}, vol.~17, no.~10, pp. 6501--6515, Oct. 2018.

\bibitem{thaj2020low}
T.~Thaj and E.~Viterbo, ``Low complexity iterative rake decision feedback
  equalizer for zero-padded {OTFS} systems,'' \emph{IEEE Trans. Veh. Technol.},
  vol.~69, no.~12, pp. 15\,606--15\,622, Dec. 2020.

\bibitem{khammammetti2018otfs}
V.~Khammammetti and S.~K. Mohammed, ``{OTFS}-based multiple-access in high
  {Doppler} and delay spread wireless channels,'' \emph{IEEE Wireless Commun.
  Lett.}, vol.~8, no.~2, pp. 528--531, Apr. 2018.

\bibitem{augustine2019interleaved}
R.~M. Augustine and A.~Chockalingam, ``Interleaved time-frequency multiple
  access using {OTFS} modulation,'' in \emph{IEEE VTC 2019}, Honolulu, HI, USA,
  Sep. 2019, pp. 1--5.

\bibitem{zhou2023otfs}
D.~Zhou, J.~Guo, S.~Wang, Z.~Zheng, Z.~Fei, W.~Yuan, and X.~Wang,
  ``{OTFS}-based robust {MMSE} precoding design in over-the-air computation,''
  \emph{IEEE Trans. Veh. Technol.}, Apr. 2024.

\bibitem{zhou2023power}
D.~Zhou, S.~Wang, J.~Guo, Z.~Zheng, W.~Wang, F.~Zesong, and X.~Wang, ``Power
  allocation for {OTFS}-based {AirComp} system with robust precoding,'' in
  \emph{Proc. IEEE ISCIT 2023}, Sydney, Australia, Oct. 2023, pp. 19--24.

\bibitem{mecklenbrauker1989tutorial}
W.~Mecklenbr{\"a}uker, ``A tutorial on non-parametric bilinear time-frequency
  signal representations,'' \emph{Time and Frequency representation of signals
  and systems}, vol. 309, pp. 11--68, 1989.

\bibitem{raviteja2018practical}
P.~Raviteja, Y.~Hong, E.~Viterbo, and E.~Biglieri, ``Practical pulse-shaping
  waveforms for reduced-cyclic-prefix {OTFS},'' \emph{IEEE Trans. Veh.
  Technol.}, vol.~68, no.~1, pp. 957--961, Jan. 2018.

\bibitem{fish2013delay}
A.~Fish, S.~Gurevich, R.~Hadani, A.~M. Sayeed, and O.~Schwartz,
  ``Delay-{Doppler} channel estimation in almost linear complexity,''
  \emph{IEEE Trans. Inf. Theory}, vol.~59, no.~11, pp. 7632--7644, Nov. 2013.

\bibitem{wei2021transmitter}
Z.~Wei, W.~Yuan, S.~Li, J.~Yuan, and D.~W.~K. Ng, ``Transmitter and receiver
  window designs for orthogonal time-frequency space modulation,'' \emph{IEEE
  Trans. Commun.}, vol.~69, no.~4, pp. 2207--2223, Apr. 2021.

\bibitem{raviteja2019embedded}
P.~Raviteja, K.~T. Phan, and Y.~Hong, ``Embedded pilot-aided channel estimation
  for {OTFS} in delay--{Doppler} channels,'' \emph{IEEE Trans. Veh. Technol.},
  vol.~68, no.~5, pp. 4906--4917, May 2019.

\bibitem{wei2022off}
Z.~Wei, W.~Yuan, S.~Li, J.~Yuan, and D.~W.~K. Ng, ``Off-grid channel estimation
  with sparse {Bayesian} learning for {OTFS} systems,'' \emph{IEEE Trans.
  Wireless Commun.}, vol.~21, no.~9, pp. 7407--7426, Sep. 2022.

\bibitem{ali1998doppler}
I.~Ali, N.~Al-Dhahir, and J.~E. Hershey, ``Doppler characterization for leo
  satellites,'' \emph{IEEE Trans. Commun.}, vol.~46, no.~3, pp. 309--313, Mar.
  1998.

\bibitem{boyd2004convex}
S.~P. Boyd and L.~Vandenberghe, \emph{Convex {Optimization}}.\hskip 1em plus
  0.5em minus 0.4em\relax U.K.: Cambridge Univ. Press, 2004.

\bibitem{strang1993introduction}
G.~Strang, \emph{Introduction to {Linear} {Algebra}}, 6th~ed.\hskip 1em plus
  0.5em minus 0.4em\relax Cambridge, MA, USA: Wellesley-Cambridge Press, 1993.

\bibitem{li2021cross}
S.~Li, W.~Yuan, Z.~Wei, and J.~Yuan, ``Cross domain iterative detection for
  orthogonal time frequency space modulation,'' \emph{IEEE Trans. Wirel.
  Commun.}, vol.~21, no.~4, pp. 2227--2242, Apr. 2021.

\end{thebibliography}
\end{document}